%% file: SMP-19-011_temp.tex
\pdfoutput=1
\documentclass[11pt,twoside,a4paper,cmspaper,final,collab]{cms-tdr}
\def\svnVersion{edea5eb-D}\def\svnDate{2021/03/02}\def\cmsCernNoTag{CERN-EP-2020-206}\def\cmsCernDate{\today}\def\cmsMessage{"Published in the Journal of High Energy Physics as \href{http://dx.doi.org/10.1007/JHEP04(2021)109}{\doi{10.1007/JHEP04(2021)109}.}"}
\begin{document}\cmsNoteHeader{SMP-19-011}

\newcommand{\zcjet}{{\PZ}+{\PQc jet}\xspace}
\newcommand{\zbjet}{{\PZ}+{\PQb jet}\xspace}
\newcommand{\zljet}{{\PZ}+{light jet}\xspace}
\newcommand{\mgamc}{MG5\_aMC\xspace}
\newcommand{\pp}{{\Pp}\Pp}
\newcommand{\ee}{{\Pe}\Pe}
\newcommand{\mumu}{{\PGm}\PGm}
\newcommand{\ptZ}{\ensuremath{\pt^{\PZ}}\xspace}
\newcommand{\ptJ}{\ensuremath{\pt^{\PQc\text{ jet}}}\xspace}
\newcommand{\ptCTJ}{\ensuremath{\pt^{{\PQc}\text{-tagged jet}}}\xspace}
\newcommand{\SVM}{\ensuremath{m_{\mathrm{SV}}}\xspace}
\newcommand{\SFl}{\ensuremath{SF_\mathrm{l}}\xspace}
\newcommand{\SFc}{\ensuremath{SF_\PQc}\xspace}
\newcommand{\SFb}{\ensuremath{SF_\PQb}\xspace}
\newcommand{\Zb}{{{\PZ} boson}\xspace}
\newcommand{\Zbs}{{{\PZ} bosons}\xspace}
\newcommand{\cjet}{{{\PQc} jet}\xspace}
\newcommand{\cjets}{{{\PQc} jets}\xspace}
\providecommand{\cmsTable}[1]{\resizebox{\textwidth}{!}{#1}}

\cmsNoteHeader{SMP-19-011} 
\title{Measurement of differential cross sections for \texorpdfstring{\PZ}{Z} bosons produced in association with charm jets in \texorpdfstring{$\Pp\Pp$}{pp} collisions at \texorpdfstring{$\sqrt{s}=13\TeV$}{sqrt(s)=13 TeV}}

\date{\today}

\abstract{
Measurements are presented of differential cross sections for the production of \Zbs in association with at least one jet 
initiated by a charm quark in \pp collisions at $\sqrt{s}=13\TeV$. 
The data recorded by the CMS experiment at the LHC correspond to an integrated luminosity of 35.9\fbinv.
The final states contain a pair of electrons or muons that are the decay products of a \Zb, 
and a jet consistent with being initiated by a charm quark produced in the hard interaction.
Differential cross sections as a function of the transverse momentum $\pt$ of the \Zb and $\pt$ of the charm jet are compared with predictions from Monte Carlo event generators.
The inclusive production cross section $405.4\pm 5.6\stat\pm 24.3\,(\text{exp}) \pm 3.7\thy\unit{pb}$,
is measured in a fiducial region requiring both leptons to have pseudorapidity $\abs{\eta}<2.4$ and $\pt>10\GeV$, 
at least one lepton with $\pt>26\GeV$, and a mass of the pair in the range 71--111\GeV, while the charm jet is required to have $\pt > 30\GeV$ and $\abs{\eta} < 2.4$.
These are the first measurements of these cross sections in proton-proton collisions at 13\TeV.
}

\hypersetup{%
pdfauthor={CMS Collaboration},%
pdftitle={Measurement of differntial cross section for Z bosons produced in association with charm jets in pp collisions at sqrt{s}=13 TeV},%
pdfsubject={CMS},%
pdfkeywords={CMS, physics, jets, charm quark}}

\maketitle 
\section{Introduction}

The CERN LHC produced a large number of events at $\sqrt{s}=13\TeV$ in proton-proton
(\pp) collisions containing a \PZ
boson accompanied by one or more jets initiated by
charm quarks (\cjets). 
The differential cross sections for inclusive \zcjet production, 
as functions of the transverse momenta \pt of the \Zb and of the \cjet, are used to verify quantum chromodynamics (QCD) models, 
provide information on the parton distribution function (PDF) of the charm quark, 
and investigate the possibility of observing the intrinsic charm quark (IC) component in the nucleon~\cite{charm1, charm2, charm3}.
An IC component would enhance the rate of \zcjet production, especially at large values of \pt of the \Zb and of the \cjet.

\begin{figure}[!htb]
\centering
\includegraphics[width=0.5\textwidth]{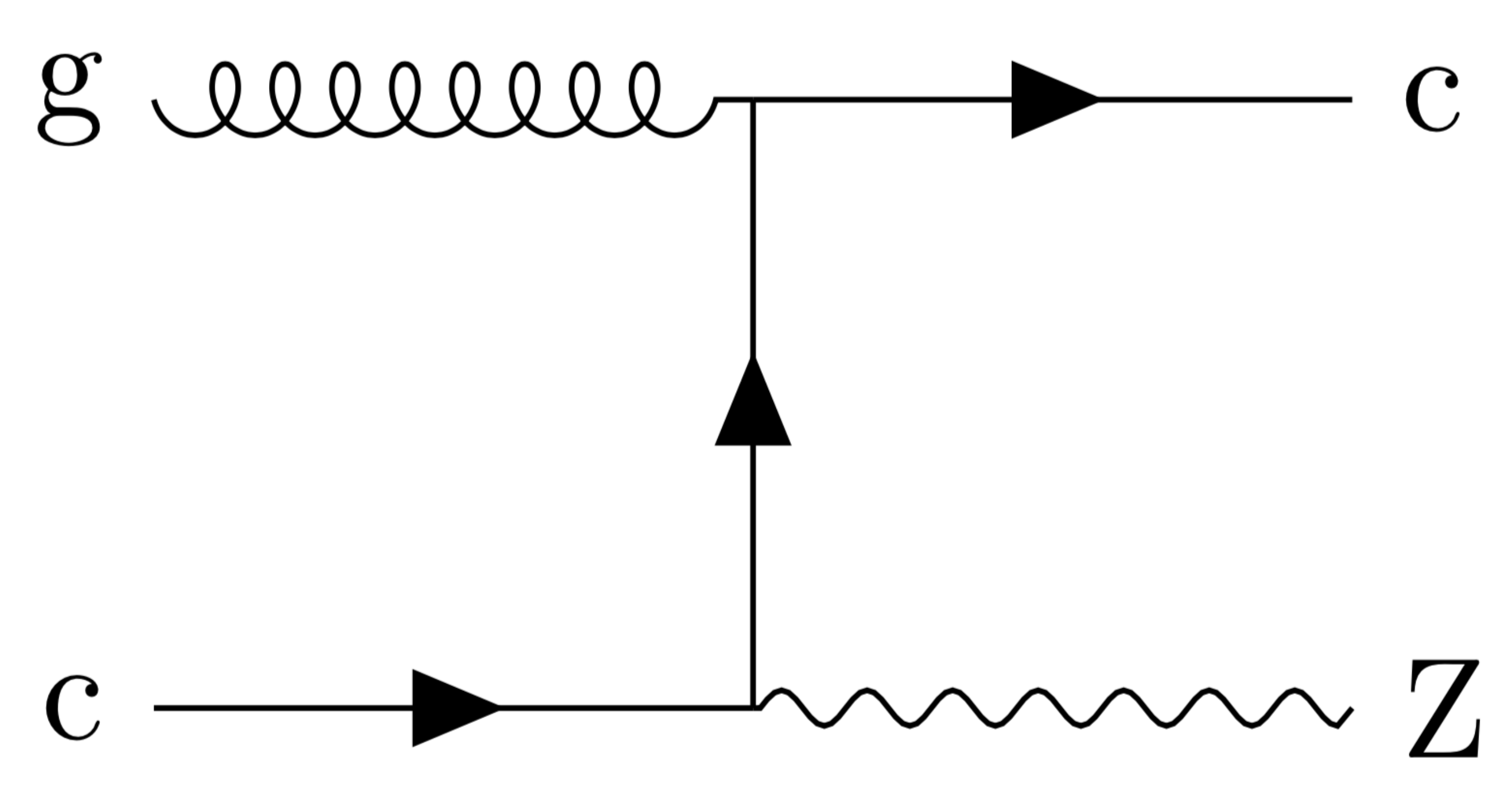}
   \caption{\label{feyn} Example Feynman diagram for the \zcjet process.}
\end{figure}

Associated production of a \Zb and a \cjet is an important background in searches for physics beyond the standard model (SM). 
For example, in supersymmetry models a top scalar quark (\PSQt) could decay into a charm quark and an undetected lightest supersymmetric particle, 
providing thereby a large \pt imbalance~\cite{cms:susy2}. One of the background sources for such a process is \zcjet production with the \Zb decaying into neutrinos. 
Better modelling of \zcjet production through studies of visible decay modes can enhance the sensitivity in searches for new physics. 
An example of a Feynman diagram corresponding to the \zcjet process is shown in Fig.~\ref{feyn}.

A previous measurement of the \zcjet cross section at 8\TeV is reported in Ref.\cite{meas8tev}. 
In this paper the \Zb is formed from an identified electron or muon pair, and the \cjet is identified by applying charm tagging criteria~\cite{ctagging}
 to reconstructed jets.
 This achieves a higher selection efficiency than in the 8\TeV measurement, where \cjets were identified by reconstructed $\PDst$(2010) mesons or low-momentum muons inside the jets.

Measurements of the fiducial total and differential cross sections of \zcjet production are presented as functions of the \pt of the \Zb and of the \cjet \pt.
 To provide a direct comparison with predictions from Monte Carlo (MC) event generators (generator level), we unfold the detector effects.

The data, corresponding to an integrated luminosity of 35.9\fbinv at $\sqrt{s} = 13\TeV$, were recorded by the CMS experiment during \pp collisions in 2016. The minimum proton bunch spacing is 25\unit{ns} with 24 interactions on average per beam crossing.

\section{The CMS detector}
The central feature of the CMS apparatus is a superconducting solenoid of 6\unit{m} internal diameter, providing a magnetic field of 3.8\unit{T}. 
A silicon pixel and strip tracker, covering a pseudorapidity region $\abs{\eta} < 2.5$, a lead tungstate crystal electromagnetic calorimeter (ECAL), and a brass and scintillator 
hadron calorimeter, with each system composed of a barrel and two endcap sections, lie within the solenoid volume. Forward calorimeters, made of steel and quartz fibers, extend $\eta$ coverage provided by the barrel and endcap detectors to $\abs{\eta} < 5$. Muons are detected in gas-ionization chambers embedded in the steel flux-return yoke outside the solenoid that cover $\abs{\eta} < 2.4$.
Events of interest are selected using a two-tiered trigger system~\cite{cms:trig}. The first level, composed of specialized hardware processors, uses information from the calorimeters and muon detectors to select events at a rate of  $\approx$100\unit{kHz} within a fixed latency of about 4\mus. The second level, known as the high-level trigger, consists of a farm of processors running full event reconstruction software optimized for fast processing, that reduces the event rate to $\approx$1\unit{kHz} before data storage.
A more detailed description of the CMS detector, together with a definition of the coordinate system and kinematic variables, can be found in Ref.~\cite{cms}.

\section{Data and simulated events}
Various MC generators are used to simulate the {\PZ}+jets background and the signal processes. 
The \MGvATNLO version 2.2.2~\cite{madgraph} (\mgamc) generator is used to simulate Drell--Yan (DY) processes,
including the \zcjet signal, calculated at next-to-leading order (NLO).
Background DY events include those with a \Zb and a jet initiated by a bottom quark (\PQb jet),
or a jet initiated by a light quark or a gluon (light jet). Samples
are made for {\PZ}+$n$~jet processes ($n\le 2$), calculated at NLO in perturbative QCD.
A second signal model is provided by using {\mgamc} to calculate leading order (LO)
matrix elements for $\pp\to \PZ + n\ \text{jets}$ ($n \le4$).
For a third signal model, \SHERPA ~v2.2 \cite{sherpa,Openloops} 
is used to generate $\pp\to \PZ + n\ \text{jets}$ events,
with $n \le 2$ at NLO and $n \le 4$ at LO.
All three signal models are normalized to the value of the inclusive \PZ + jets cross section calculated at
next-to-next-to-leading order with \FEWZ v3.1~\cite{fewz}.
These samples are generated using the NNPDF 3.0~\cite{nnpdf30} PDF set.

In addition to events with light and \PQb jets, there are contributions to the background from
processes producing top quark pairs~\cite{ttbar1,ttbar2} and single top quarks~\cite{stop,stop2}.
These samples are generated using NLO \POWHEG
v2.0~\cite{poweg1,poweg2,poweg3} or \mgamc.
There is also background from vector boson pair production, which is simulated using \PYTHIA8~ v8.212~\cite{pythia8}. 

All samples, except \SHERPA, use \PYTHIA8 to model the initial- and final-state parton showers and hadronization,
with the CUETP8M1~\cite{cuetp8m1} or CUETP8M2T4~\cite{cuetp8m2t4} (top pair sample) tune that includes the NNPDF 2.3
\cite{nnpdf23} LO PDFs and the value of the strong coupling at the
mass of the \Zb is $\alpS(m_{\PZ}) =  0.119$.
Matching between the matrix element generators and the parton shower is done using the \kt--MLM~\cite{ktmlm1,ktmlm2}
scheme with the matching scale set at 19\GeV for the LO {\mgamc} samples, and the FxFx~\cite{fxfx} scheme
with the matching scale set to 30\GeV for the NLO {\mgamc} events.

\GEANTfour~\cite{geant} is used for CMS detector simulation.
The simulation includes additional $\pp$ interactions (pileup) in the current and nearby bunch crossings. 

The simulated events are reconstructed with the same algorithms used for the data.
 
\section{Object reconstruction and event selection}
The particle flow (PF) algorithm~\cite{pf} is used to reconstruct and identify individual particle candidates (physics objects) in an event,
through an optimized combination of information from the various elements of the CMS detector.
Energy depositions are measured in the calorimeters, and charged particles are identified in the
central tracking and muon systems.

Electrons are reconstructed from tracks, fitted with a Gaussian sum filter, matching energy deposits in the ECAL~\cite{electrons}.
Identification requirements are applied based on the ECAL shower shape, matching between
the track and the ECAL deposits, and observables characterizing the emission
of bremsstrahlung radiation along the electron trajectory. Electrons are required to
originate from the primary vertex, which is the vertex candidate with the largest value of summed physics-object ${\pt}^2$. 
Longitudinal and transverse impact parameters with respect to the primary vertex for barrel (endcap) are required to be $<$0.10\,(0.20) and $<$0.05\,(0.10)\cm, respectively.
 The electron momentum is estimated by combining the energy
measurement in the ECAL with the momentum measurement in the tracker. The momentum resolution for electrons with
$\pt \approx 45\GeV$ from $\PZ\to\ee$ decays ranges from 1.7 to 4.5\%. The dielectron mass resolution
for $\PZ\to\ee$ decays when both electrons are in the ECAL barrel is 1.9\%, degrading to
2.9\% when both electrons are in the endcaps.

Muons are reconstructed by combining signals from the inner tracker and the muon detector subsystems.
They are required to satisfy standard identification criteria based on the number of hits in each detector,
the track fit quality, and the consistency with the primary vertex by requiring the longitudinal and transverse
impact parameters to be less than 0.5 and 0.2\cm, respectively.
The efficiency to reconstruct and identify muons is greater than 96\%~\cite{muons}.
Matching muons to tracks measured in the silicon tracker results in a relative \pt resolution for muons with $20 < \pt <100\GeV$ of 1\% in the barrel and  3\% in the endcaps.

To reduce the misidentification rate, electrons and muons are required to be isolated.
The isolation of electron or muon is defined as the sum of the \pt of all additional PF candidates within a cone of radius
$\Delta R = \sqrt{\smash[b]{(\Delta \eta)^2 + (\Delta \phi)^2}} = 0.3\,(0.4)$  
around the electron (muon) track, where $\phi$ is the azimuthal angle in radians. After compensating for the contribution from pile{\-}up~\cite{pucorr},
the resultant sum is required to be less than 25\% of the lepton \pt.

Jets are clustered from PF candidates using the infrared- and collinear-safe anti-\kt algorithm
with a distance parameter of 0.4, as implemented in the \FASTJET package \cite{akt,fjet}.
The jet momentum is determined as the vectorial sum of all particle momenta in the jet,
and, based on simulation, is within 5 to 10\% of the true momentum over the entire \pt spectrum and detector acceptance.
To mitigate the effects of pileup, charged particle candidates identified as originating from pileup vertices are discarded and a
correction~\cite{pucorr} is applied to remove remaining contributions.
The reconstructed jet energy scale (JES) is corrected using a factorized model to
compensate for the nonlinear and nonuniform response in the calorimeters. 
Corrections are derived from simulation to bring the
measured response of jets to that of generator-level jets on
average. In situ measurements of the momentum balance in dijet,
multijet, photon+jet, and leptonically decaying {\PZ}+jet events are
used to correct any residual differences between the JES
 in data and simulation~\cite{cms:jes}.
The jet energy in simulation is degraded to match the resolution observed in data. 
The jet energy resolution (JER) amounts typically to 15\% at 10\GeV, 8\% at 100\GeV, and
4\% at 1\TeV.
Additional selection criteria are applied to remove jets potentially dominated by anomalous contributions from various sub{\-}detector components or reconstruction failures~\cite{etmiss}.
Jets identified as likely coming from pileup~\cite{pileup_2020} are also removed.

Events are selected online through a single electron trigger requiring at least one
electron candidate with $\pt> 27\GeV$ (electron channel), or a single muon trigger requiring at least one muon candidate
with $\pt>24\GeV$ (muon channel).
Offline, we require a pair of oppositely charged electrons or muons each satisfying identification and isolation criteria, 
with $\pt >  10\GeV$ and $\abs{\eta} < 2.4$, and with an invariant mass close to the
mass of the \Zb: $71 < m_{\ee~\text{or}~\mumu} < 111\GeV$.
To exceed the trigger threshold in the electron channel at least one electron must have $\pt > 29\GeV$, and in the muon channel at least one muon must have $\pt >  26\GeV$.
Small residual differences in the trigger, identification, and isolation efficiencies between data and simulation are
measured using tag-and-probe methods~\cite{wz}, and corrected by applying scale factors to simulated events.

The event must contain at least one jet with $\pt>30\GeV$ and $\abs{\eta}
<2.4$, satisfying
tight {\PQc} tagging criteria using the deep combined secondary vertices algorithm~\cite{ctagging}. This algorithm 
discriminates \cjets from \PQb and light jets based on jet properties
such as the presence of displaced vertices (secondary vertices) and tracks with large impact parameter with respect to the primary vertex.
Data from {\PW}+jets, \ttbar, and inclusive jet production are used
to measure the \PQc tagging efficiency for \cjets, and
mistag rates for \PQb and light jets. These are
compared with the simulation, where the reconstructed jet flavor is
known from its hadron content.
Small differences between data and simulation are corrected by
applying scale factors to the simulation.
The threshold applied in this analysis
gives a \PQc tagging efficiency of about 30\%, and misidentification
probabilities of 1.2\% for light jets and 20\% for \PQb jets, 
with relative uncertainties between 5\% and 15\% depending on the \pt of the jet.
If several {\PQc}-tagged jets occur in the event, the one with the
highest $\pt$ is selected.

The simulated events are classified according to generator-level information.
Generator-level jets are made by clustering
all stable particles resulting from hadronization using the anti-\kt algorithm
with a distance parameter of 0.4, and the jet flavor is defined by the
flavor of the hadrons within the jet. If an event contains a
generator-level jet with $\pt > 10\GeV$ containing a \PQb hadron,
the event is defined as a \zbjet event.
If there is no such generator-level \PQb jet in the event and there
is at least one generator-level jet with $\pt > 10\GeV$
containing a \PQc hadron,
the event is defined as a \zcjet event. Other events in the DY sample are classified as \zljet events.
The generator-level leptons are dressed by adding the momenta of all photons within $\Delta R  = 0.1 $ around the lepton directions.

\section{Signal determination and data unfolding}
Measurement of the differential cross sections of \zcjet
production as a function of the \pt of the \Zb (\ptZ) and as
a function of the \cjet \pt (\ptJ) are performed in several steps.
The first step is to select \cjet-enriched samples of $\PZ\to\Pe\Pe$
(electron channel) or $\PZ\to\PGm\PGm$ (muon channel) candidate events.
The second step is to split the sample into different bins according
to the \pt of the \Zb or {\PQc}-tagged jet (\ptCTJ), and to measure the
number of \zcjet events in each bin.
The third step is to unfold the data, using the simulation of the signal to construct response matrices to relate the observed
distributions to those at the generator level.
The final step is to combine the resulting unfolded electron and muon
channel \pt distributions, and compare them with predictions from
different MC event generators.

\begin{figure}[!htb]
    \centering
    \includegraphics[width=0.49\textwidth]{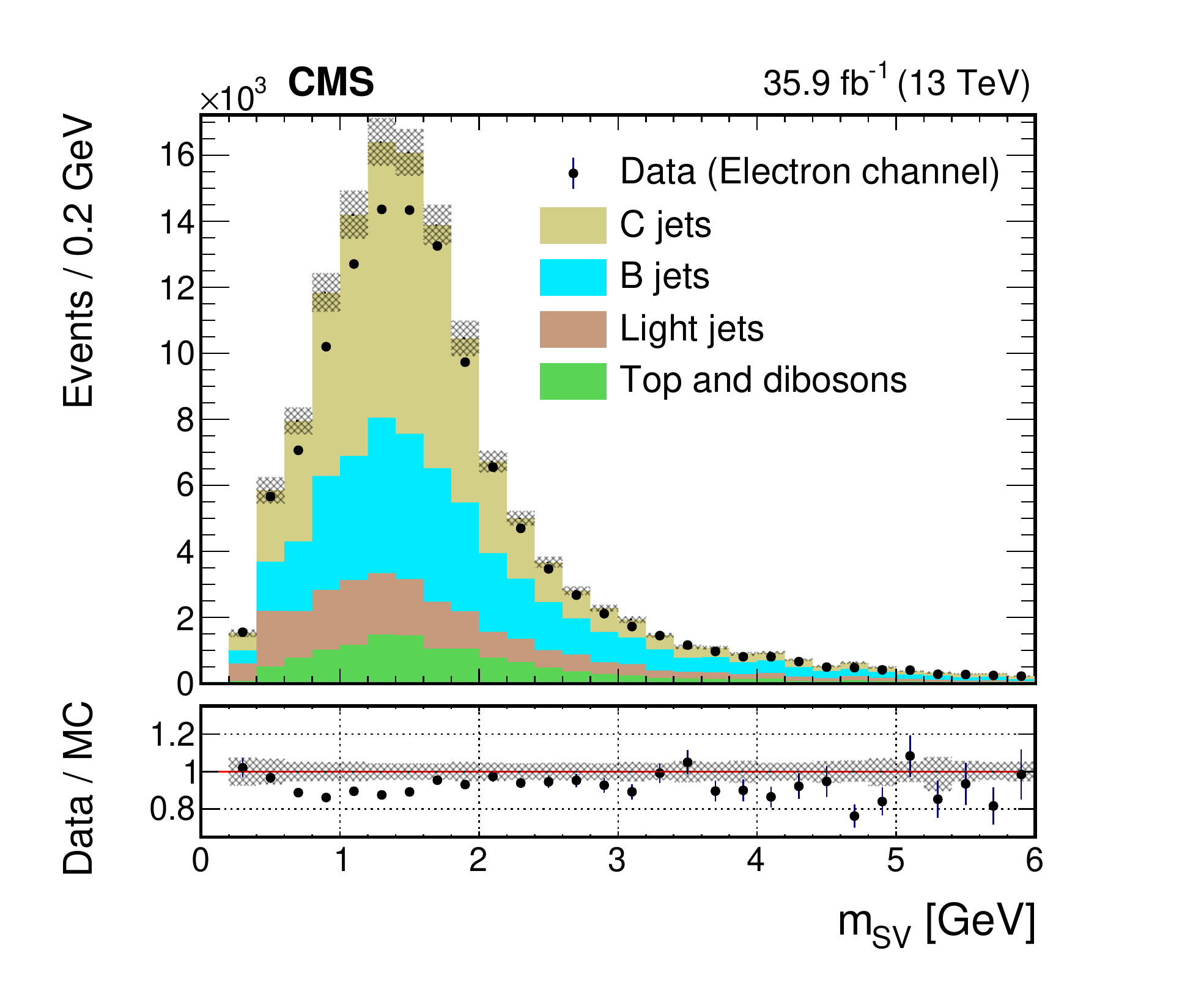}
    \includegraphics[width=0.49\textwidth]{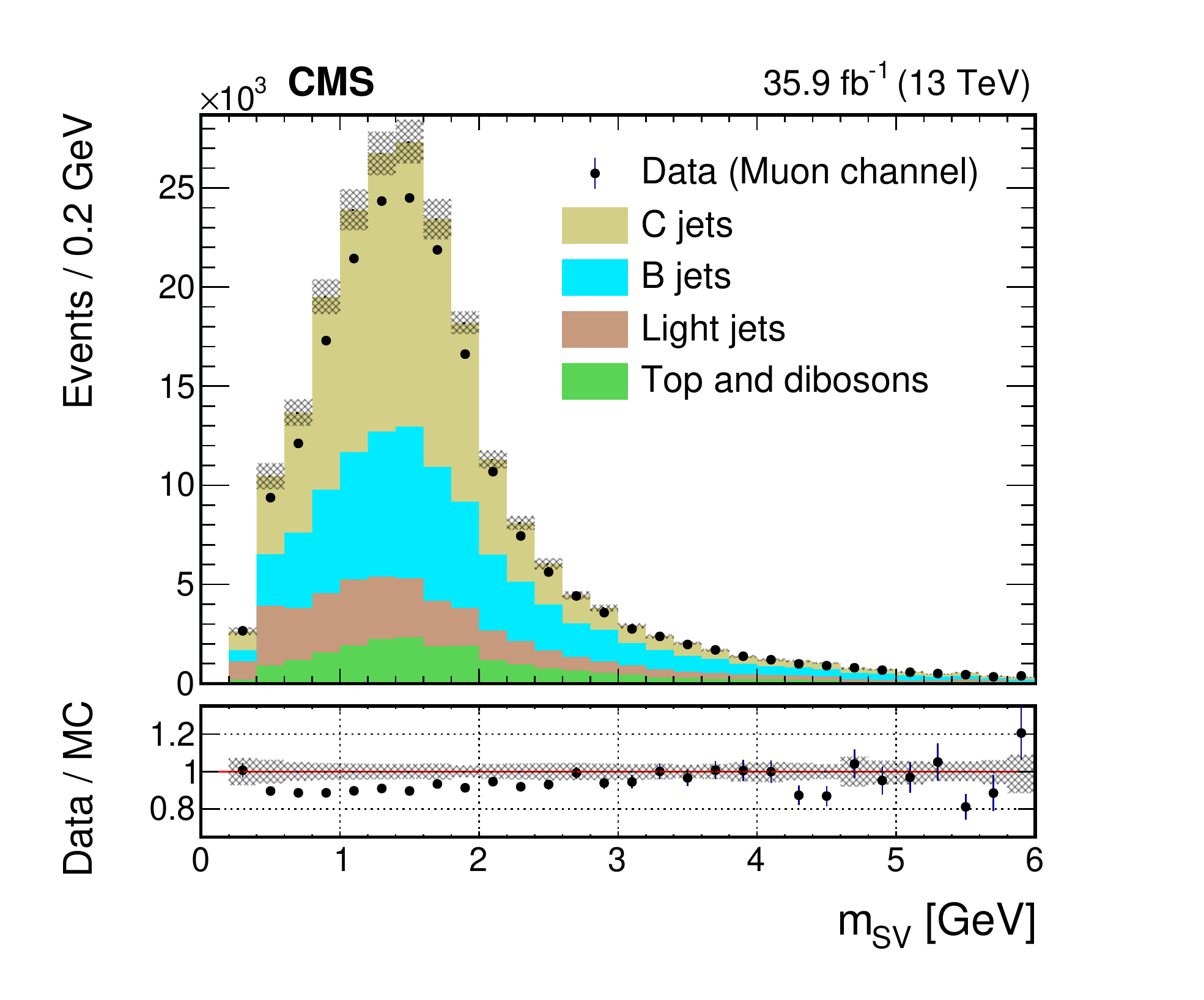} 
    \caption{ \label{ctag:SVM} Distribution of the secondary vertex
      mass \SVM of the highest-{\pt}  {\PQc}-tagged central jet, for electron
      (left) and muon (right) channels. The observed data is
      compared with the different signal and background components in
      simulation, before normalization scale factors are applied. Dashed area represents MC systematic uncertainties. The vertical bars on the data points represent statistical uncertainties. Bottom panels on each plot represent the data to MC ratio.}
\end{figure}

\begin{figure}[!htb]
    \centering
    \includegraphics[width=0.49\textwidth]{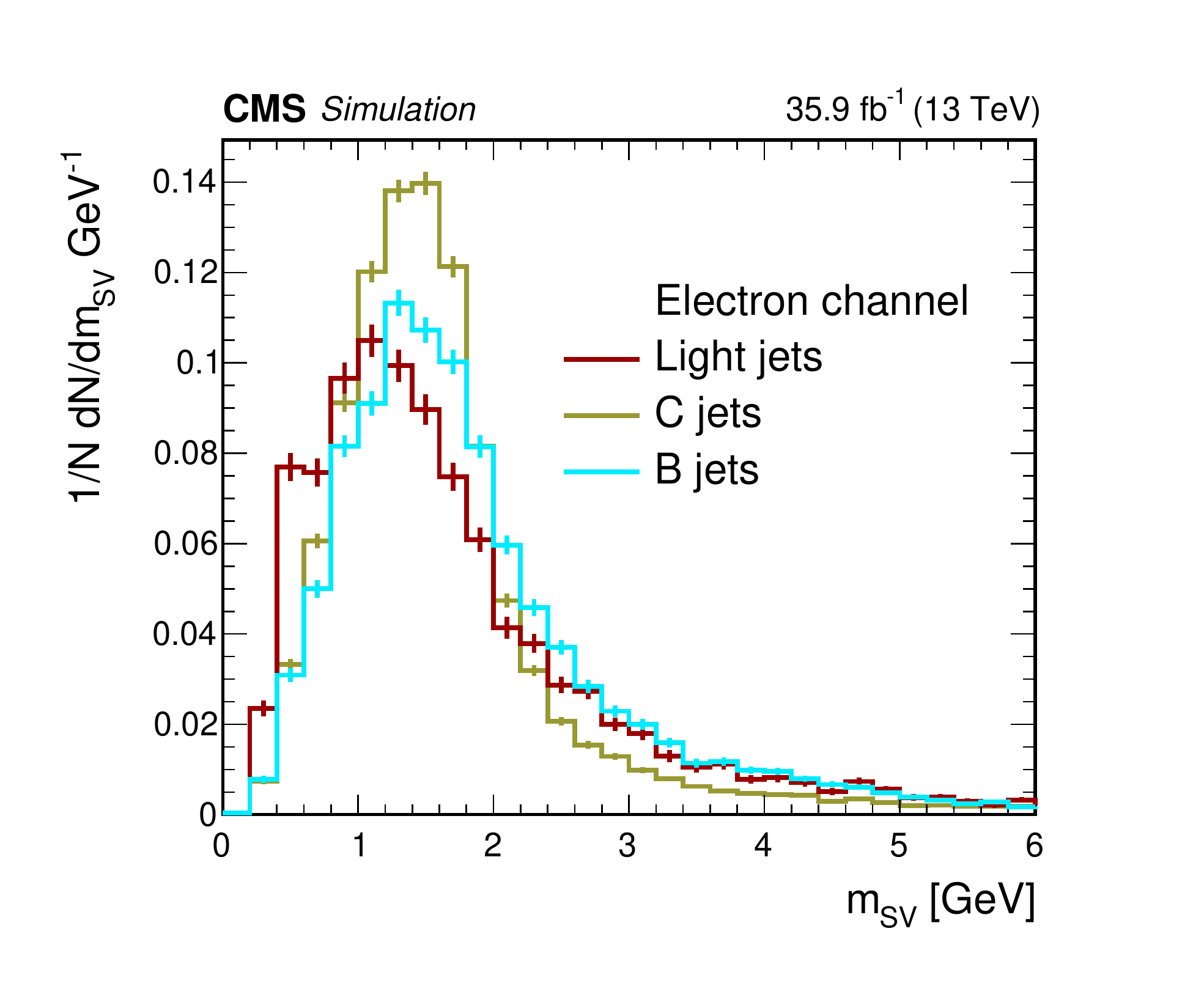}
    \includegraphics[width=0.49\textwidth]{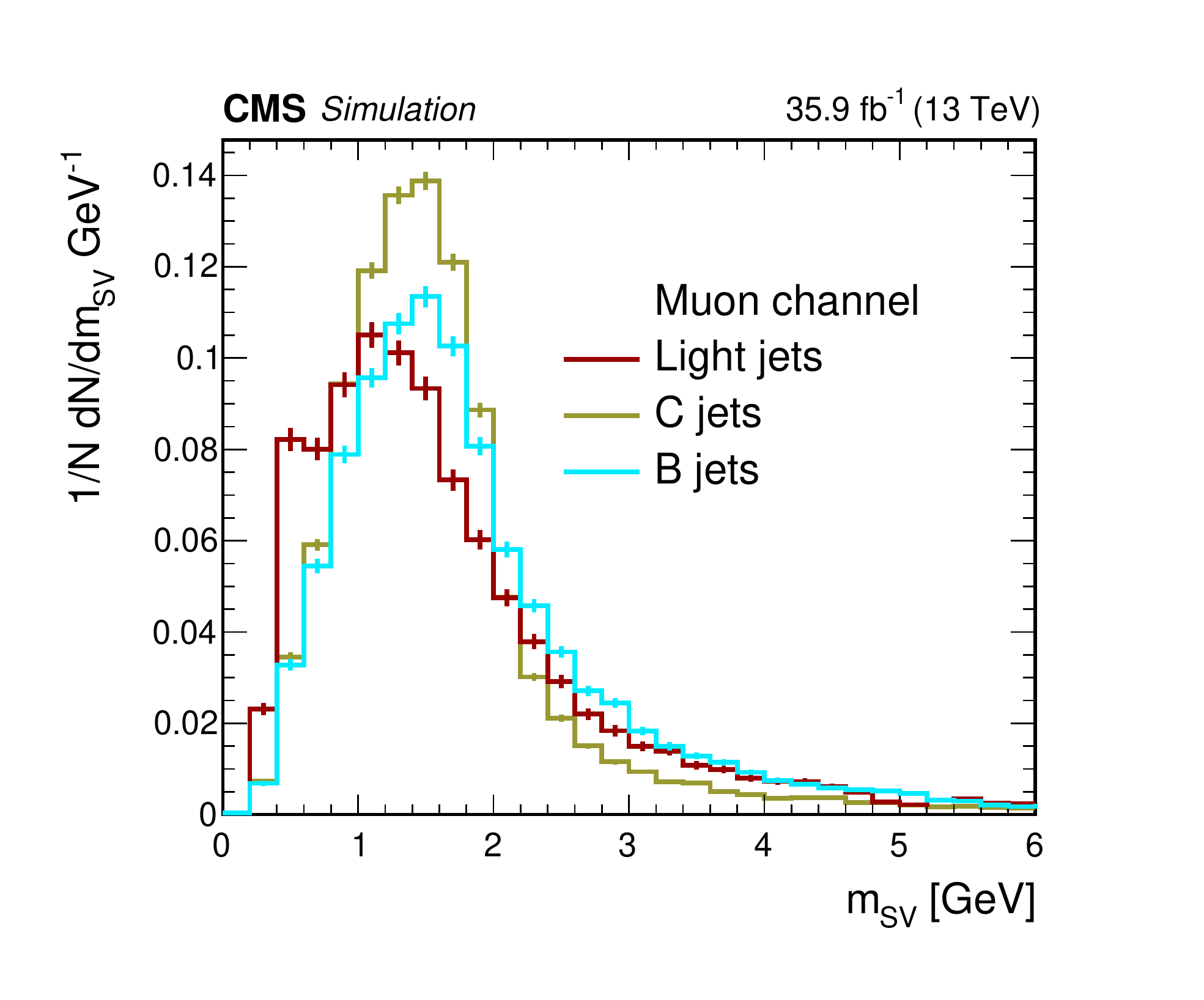}
    \caption{ \label{shape} Distribution of the secondary vertex
      mass of the highest-{\pt}  {\PQc}-tagged central jet, for electron
      (left) and muon (right) channels for \zljet, \zcjet and \zbjet components, normalized to 1. Vertical bars represent statistical uncertainties.}
\end{figure}

Charm hadrons can decay at points displaced from the primary vertex.
This secondary vertex is reconstructed using the inclusive vertex
finder algorithm~\cite{IVF}.
The invariant mass of charged particles associated with the secondary vertex
(\SVM) in the {\PQc}-tagged jet~\cite{ctagging} is used to discriminate between
signal and background.
Figure~\ref{ctag:SVM} shows the observed distributions of \SVM in the electron and muon channels, 
compared with the different signal and background contributions
predicted by the simulation.
Although \SVM is an ingredient in the \PQc tagging algorithm, 
there are sufficient differences remaining in the distributions for the {\PQc}-tagged samples to provide information on the flavor composition.
The normalized distributions of \SVM for \zljet, \zcjet and \zbjet components are compared in Fig.~\ref{shape}.

The top quark and diboson background predictions are taken
directly from simulation. 
The  normalizations for the \zcjet, \zbjet and \zljet components are
then obtained by fitting templates of the \SVM
distribution obtained from simulation to the observed data. 
A maximum likelihood template fit is performed separately in
each bin of \Zb candidate or {\PQc}-tagged jet \pt.

\begin{figure}[!ht]
    \centering
   \includegraphics[width=0.49\textwidth]{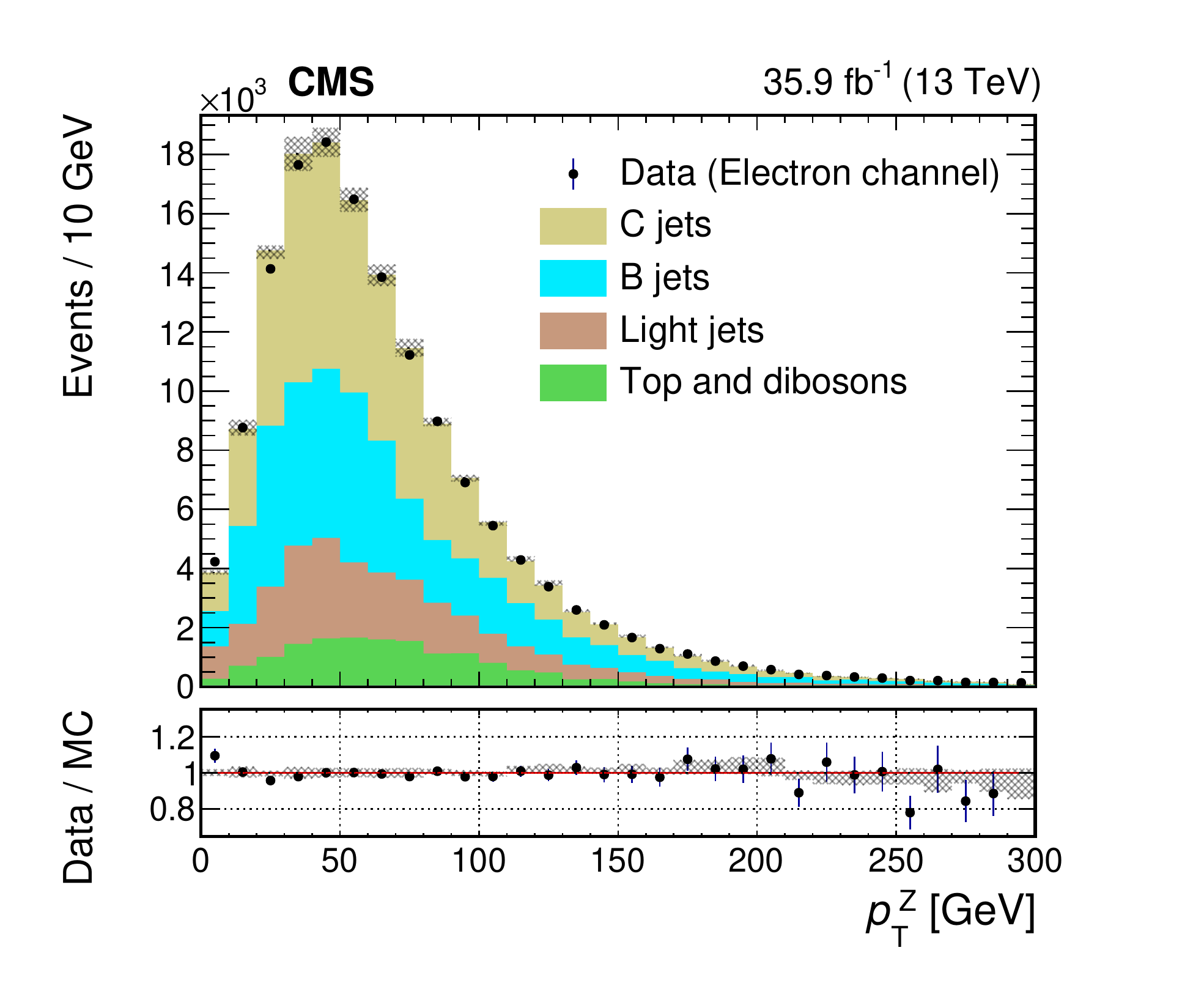}
\includegraphics[width=0.49\textwidth]{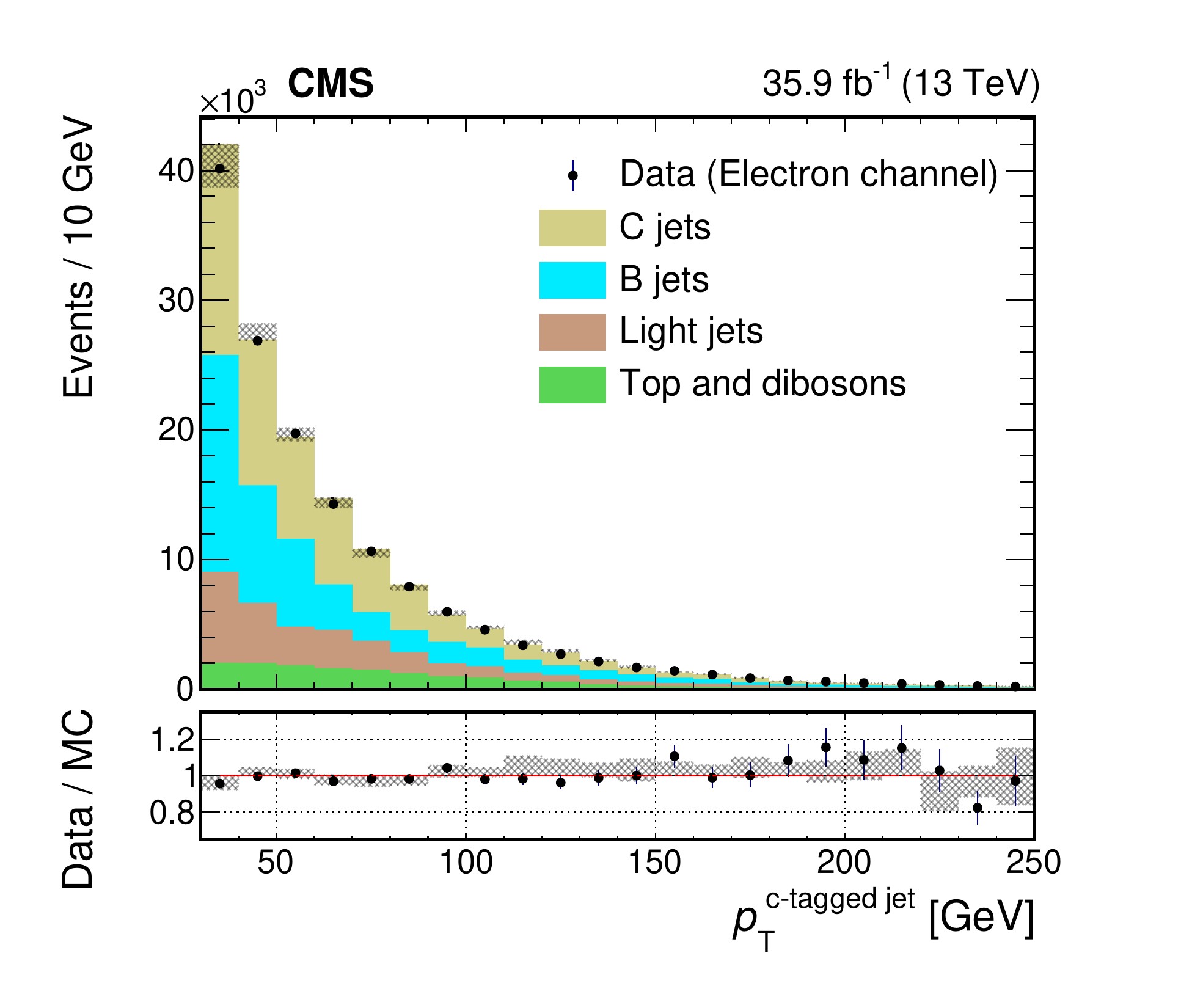} \\
    \includegraphics[width=0.49\textwidth]{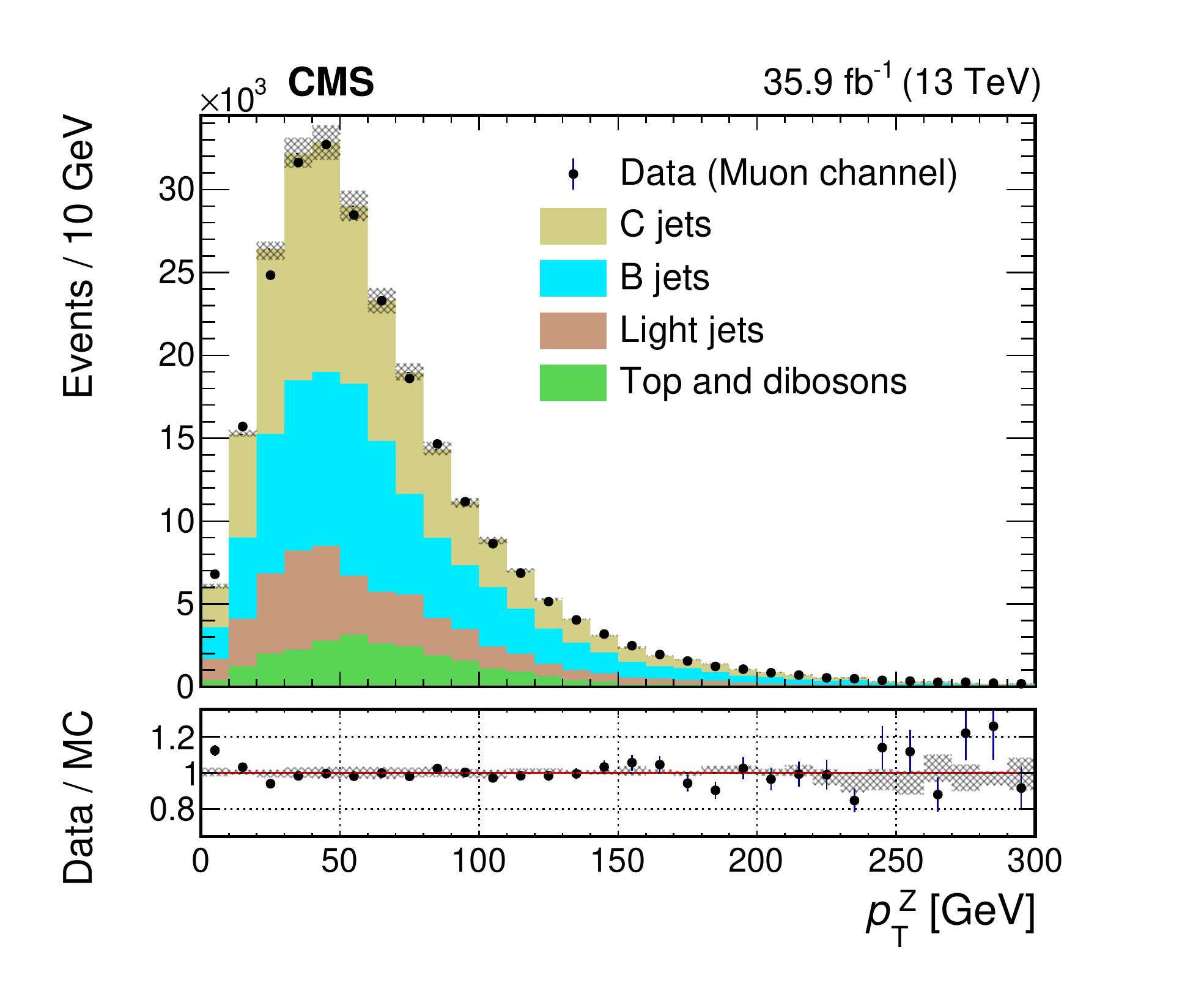} 
\includegraphics[width=0.49\textwidth]{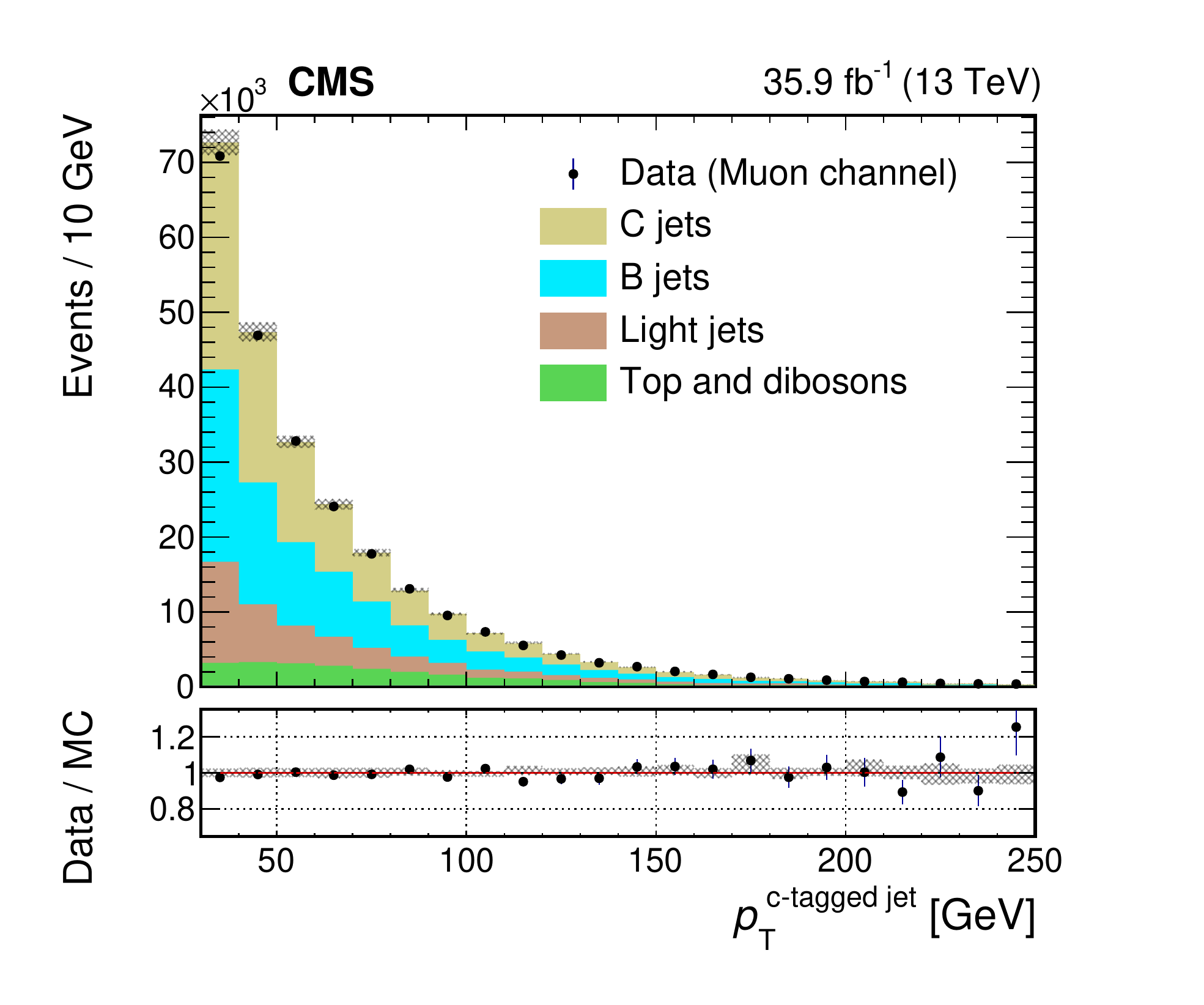}
    \caption{\label{kfactors:test} The distributions of \pt in data
      and corrected simulation,  after applying the fitted scale
      factors to the Drell-Yan components. The upper plots show
      distributions for the electron channel, with the \pt of the electron
      pair (left) and {\PQc}-tagged jet (right). The lower plots show
      distributions for the muon channel with the \pt of the muon pair (left) and {\PQc}-tagged jet (right). Dashed area represents MC systematic uncertainties. The vertical bars on the data points represent statistical uncertainties. Bottom panels on each plot represent the data to MC ratio.}
\end{figure}

The values of the scale factors for the light (\SFl), charm (\SFc), and bottom (\SFb) components, defined as the ratio of the fitted
normalization to the prediction from simulation, are presented in
Tables~\ref{kfactors:JE}--\ref{kfactors:Z} for each \pt bin for each {{\PZ}+{\cPq} jet\xspace}
process. 
The correlation coefficients between errors of different flavor SFs, 
found in the fit, are approximately equal to $-0.8$, $-0.35$ and $-0.25$ for \SFc and \SFb, \SFl and \SFb, \SFc and \SFl respectively.
Sources of systematic uncertainty are discussed in Sec.~\ref{sec:syst}.
Figure~\ref{kfactors:test} shows the distributions of the \Zb candidate
and {\PQc}-tagged jet \pt after applying these scale factors,
assuming they are constant across the \pt range in which they are
determined.
The post-fit \SVM distributions are presented in Appendix~\ref{app:svm}.
Good agreement is observed between simulation and data after applying these factors. 

\begin{table}[!htb]
\centering
\renewcommand{\arraystretch}{1.3}
\topcaption{Values of  \zljet \SFl, \zcjet \SFc, and \zbjet \SFb scale factors measured in the electron channel, as
  a function of {\PQc}-tagged jet \pt.
  The first uncertainty is the statistical uncertainty from the fit, the second is the systematic uncertainty.}
\begin{tabular}{cccc}
{\PQc}-tagged jet \pt (\GeVns)& \SFl & \SFc & \SFb\\
\hline
30--37 & 1.16 $\pm$ 0.05 $ ^{+0.29} _{-0.21}$ & 0.70 $\pm$ 0.04 $^{+0.09} _{-0.11}$ & 1.06 $\pm$ 0.05 $^{+0.16} _{-0.12}$ \\ 
37--45 & 0.79 $\pm$ 0.06 $ ^{+0.26} _{-0.18}$ & 0.89 $\pm$ 0.03 $^{+0.10} _{-0.08}$ & 0.92 $\pm$ 0.05 $^{+0.16} _{-0.16}$ \\ 
45--60 & 0.97 $\pm$ 0.06 $ ^{+0.23} _{-0.19}$ & 0.74 $\pm$ 0.03 $^{+0.08} _{-0.06}$ & 1.07 $\pm$ 0.06 $^{+0.09} _{-0.09}$ \\ 
60--90 & 0.99 $\pm$ 0.07 $ ^{+0.24} _{-0.18}$ & 0.87 $\pm$ 0.04 $^{+0.08} _{-0.08}$ & 0.95 $\pm$ 0.07 $^{+0.09} _{-0.12}$ \\ 
90--250 & 0.92 $\pm$ 0.07 $ ^{+0.27} _{-0.19}$ & 0.98 $\pm$ 0.05 $^{+0.11} _{-0.10}$ & 1.04 $\pm$ 0.07 $^{+0.08} _{-0.08}$ \\ 
\end{tabular}
\label{kfactors:JE}
\end{table}

\begin{table}[h]
\centering
\renewcommand{\arraystretch}{1.3}
\topcaption{Values of \zljet \SFl, \zcjet \SFc, and \zbjet \SFb scale factors measured in the electron channel, as
  a function of \PZ candidate \pt.
  The first uncertainty is the statistical uncertainty from the fit, the second is the systematic uncertainty.}
\begin{tabular}{cccc}
\PZ candidate \pt (\GeVns)& \SFl & \SFc & \SFb\\
\hline
0--30 & 0.86 $\pm$ 0.05 $ ^{+0.24} _{-0.18}$ & 0.76 $\pm$ 0.04 $^{+0.11} _{-0.09}$ & 1.25 $\pm$ 0.07 $^{+0.17} _{-0.18}$ \\ 
30--50 & 0.98 $\pm$ 0.05 $ ^{+0.23} _{-0.17}$ & 0.80 $\pm$ 0.03 $^{+0.08} _{-0.08}$ & 0.91 $\pm$ 0.05 $^{+0.10} _{-0.08}$ \\ 
50--65 & 0.85 $\pm$ 0.07 $ ^{+0.21} _{-0.16}$ & 0.78 $\pm$ 0.04 $^{+0.15} _{-0.11}$ & 1.04 $\pm$ 0.06 $^{+0.17} _{-0.17}$ \\ 
65--95 & 1.14 $\pm$ 0.08 $ ^{+0.30} _{-0.22}$ & 0.97 $\pm$ 0.04 $^{+0.08} _{-0.07}$ & 0.76 $\pm$ 0.06 $^{+0.09} _{-0.11}$ \\ 
95--300 & 1.01 $\pm$ 0.07 $ ^{+0.26} _{-0.20}$ & 0.83 $\pm$ 0.05 $^{+0.07} _{-0.07}$ & 1.13 $\pm$ 0.07 $^{+0.08} _{-0.08}$ \\ 
\end{tabular}
\label{kfactors:ZE}
\end{table}

\begin{table}[ht]
\centering
\renewcommand{\arraystretch}{1.3}
\topcaption{Values of \zljet \SFl, \zcjet \SFc, and \zbjet \SFb scale factors measured in the muon channel, as
  a function of {\PQc}-tagged jet \pt.
  The first uncertainty is the statistical uncertainty from the fit, the second is the systematic uncertainty.}
\begin{tabular}{cccc}
{\PQc}-tagged jet \pt (\GeVns)& \SFl &  \SFc & \SFb\\
\hline
30--37 & 0.95 $\pm$ 0.04 $ ^{+0.24} _{-0.17}$ & 0.82 $\pm$ 0.03 $^{+0.12} _{-0.07}$ & 1.04 $\pm$ 0.05 $^{+0.11} _{-0.19}$ \\ 
37--45 & 0.93 $\pm$ 0.05 $ ^{+0.26} _{-0.23}$ & 0.82 $\pm$ 0.03 $^{+0.06} _{-0.06}$ & 0.96 $\pm$ 0.05 $^{+0.12} _{-0.09}$ \\ 
45--60 & 0.81 $\pm$ 0.04 $ ^{+0.20} _{-0.15}$ & 0.79 $\pm$ 0.03 $^{+0.09} _{-0.06}$ & 1.10 $\pm$ 0.04 $^{+0.08} _{-0.07}$ \\ 
60--90 & 0.88 $\pm$ 0.04 $ ^{+0.23} _{-0.17}$ & 0.80 $\pm$ 0.03 $^{+0.06} _{-0.08}$ & 1.25 $\pm$ 0.05 $^{+0.12} _{-0.10}$ \\ 
90--250 & 0.92 $\pm$ 0.05 $ ^{+0.24} _{-0.17}$ & 0.79 $\pm$ 0.04 $^{+0.07} _{-0.06}$ & 1.16 $\pm$ 0.06 $^{+0.12} _{-0.12}$ \\ 
\end{tabular}
\label{kfactors:J}
\end{table}

\begin{table}[h]
\centering
\renewcommand{\arraystretch}{1.3}
\topcaption{Values of \zljet \SFl, \zcjet \SFc, and \zbjet \SFb scale factors measured in the muon channel, as
  a function of \PZ candidate \pt.
  The first uncertainty is the statistical uncertainty from the fit, the second is the systematic uncertainty.}
\begin{tabular}{cccc}
\PZ candidate \pt (\GeVns)& \SFl & \SFc & \SFb\\
\hline
0--30 & 0.97 $\pm$ 0.04 $ ^{+0.24} _{-0.20}$ & 0.82 $\pm$ 0.03 $^{+0.09} _{-0.08}$ & 1.09 $\pm$ 0.05 $^{+0.11} _{-0.10}$ \\ 
30--50 & 0.91 $\pm$ 0.04 $ ^{+0.21} _{-0.16}$ & 0.80 $\pm$ 0.02 $^{+0.07} _{-0.06}$ & 0.99 $\pm$ 0.04 $^{+0.05} _{-0.06}$ \\ 
50--65 & 0.63 $\pm$ 0.06 $ ^{+0.17} _{-0.13}$ & 0.73 $\pm$ 0.03 $^{+0.09} _{-0.06}$ & 1.24 $\pm$ 0.05 $^{+0.09} _{-0.10}$ \\ 
65--95 & 0.96 $\pm$ 0.05 $ ^{+0.25} _{-0.18}$ & 0.85 $\pm$ 0.03 $^{+0.09} _{-0.06}$ & 1.04 $\pm$ 0.05 $^{+0.13} _{-0.14}$ \\ 
95--300 & 0.89 $\pm$ 0.05 $ ^{+0.23} _{-0.17}$ & 0.78 $\pm$ 0.04 $^{+0.07} _{-0.07}$ & 1.33 $\pm$ 0.06 $^{+0.08} _{-0.08}$ \\ 
\end{tabular}
\label{kfactors:Z}
\end{table}

The generator-level signal is defined to be \zcjet events with
two oppositely charged generator-level electrons or muons
with $\pt > 10\GeV$ (at least one with $\pt > 26\GeV$), $\abs{\eta} < 2.4$,  and an invariant mass $71 < m_{\ee~\text{or}~\mumu} < 111\GeV$. 
There must also be at least one generator-level \cjet with $\pt >
30\GeV$ and $\abs{\eta} < 2.4$.
To avoid double counting, jets within $\Delta R = 0.4$ of one of the
two leptons from the \PZ candidate are removed.

\begin{figure}[!htb]
    \centering
    \includegraphics[width=0.49\textwidth]{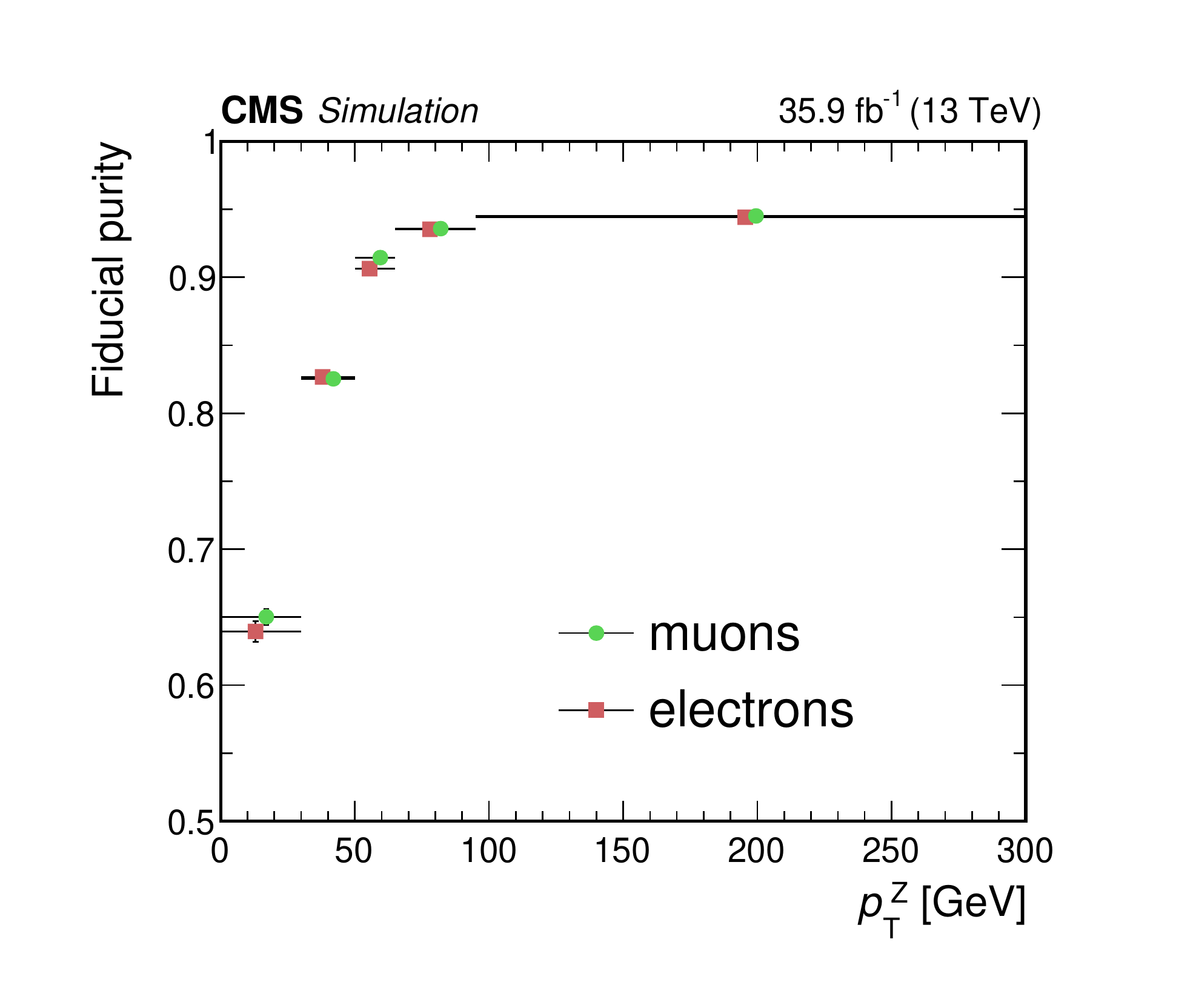}
\includegraphics[width=0.49\textwidth]{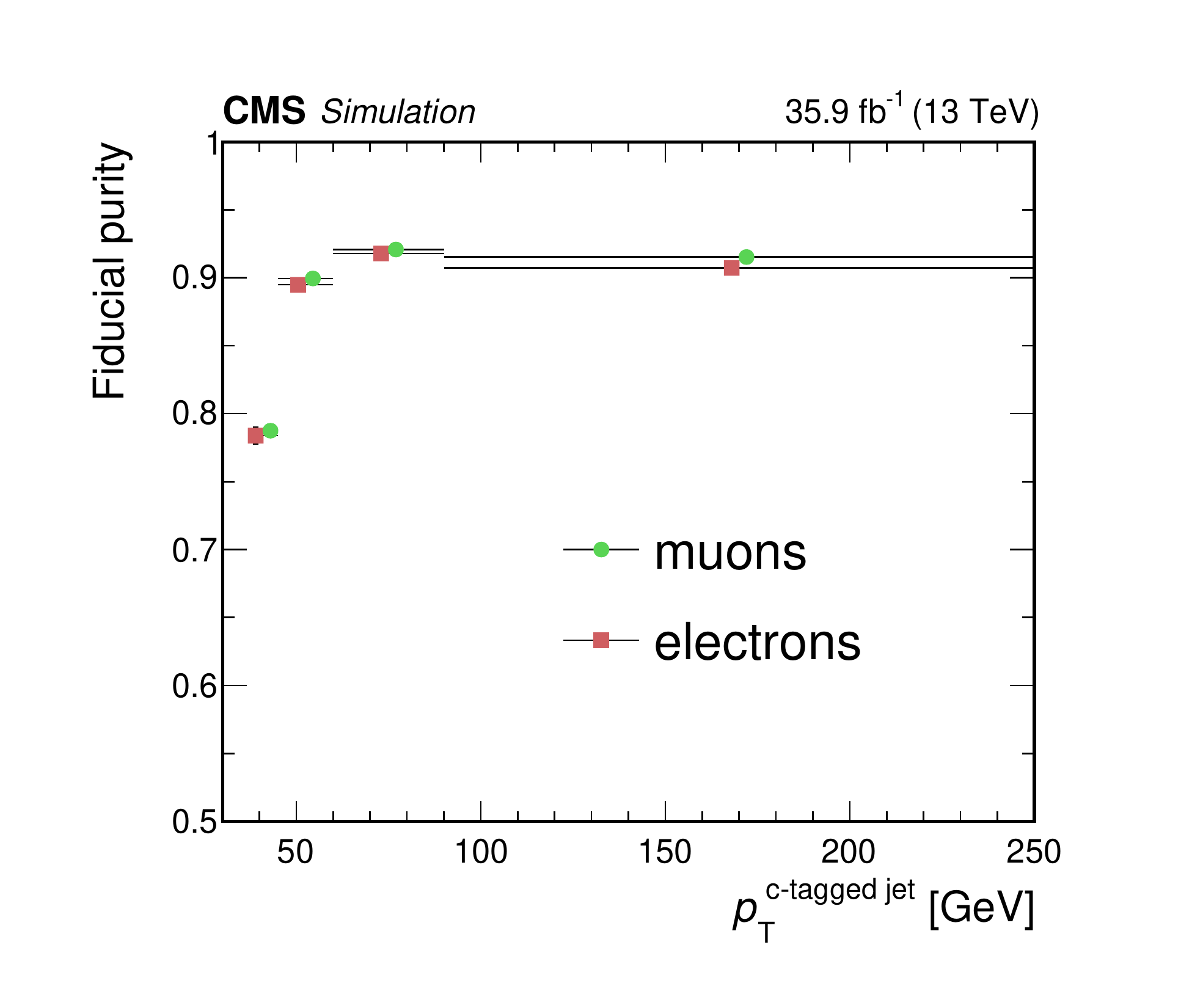} 
\caption{\label{unf:bkg}Fraction of selected \zcjet events originating
  within the fiducial phase space as a function of \pt. The plots show distributions for electron and muons channels
  as a function of \ptZ (left) and \ptCTJ (right). 
}
\end{figure}

A fraction of \zcjet events that are outside the signal phase space
will migrate into the reconstructed signal region, primarily events
with \cjets with generated $\pt < 30\gev$ but reconstructed
$\pt >30\gev$ due to the finite detector resolution. The fraction of \zcjet events that are inside the signal phase space is estimated
from the number of selected events in which the {\PQc}-tagged jet and lepton pair match within $\Delta R < 0.3$ to a generator-level highest-\pt
~\cjet and lepton pair satisfying the phase space requirements.
Figure~\ref{unf:bkg} shows this fraction as functions of \Zb and {\PQc}-tagged jet \pt,
for electron and muon channels, calculated using \MGvATNLO sample.

Response matrices are constructed using the \zcjet events in the DY
sample that is simulated using the \mgamc (NLO) generator, and cross-checked
using the \mgamc (LO) generator. Each matrix
entry represents the probability for an event generated in the signal phase space
within a certain \cjet (or \Zb) \pt range to end up 
within a certain reconstructed \cjet (or \Zb candidate) \pt range.
The unfolding was done with 5 detector-level \pt bins and 4 generator-level \pt bins. 
The \textsc{TUnfold} package v17.5~\cite{unfold}, which is based on a least-squares
fit, is then used to invert the response matrices and unfold the
distribution of the measured number of \zcjet events.

\begin{figure}[!htb]
    \centering
    \includegraphics[width=0.49\textwidth]{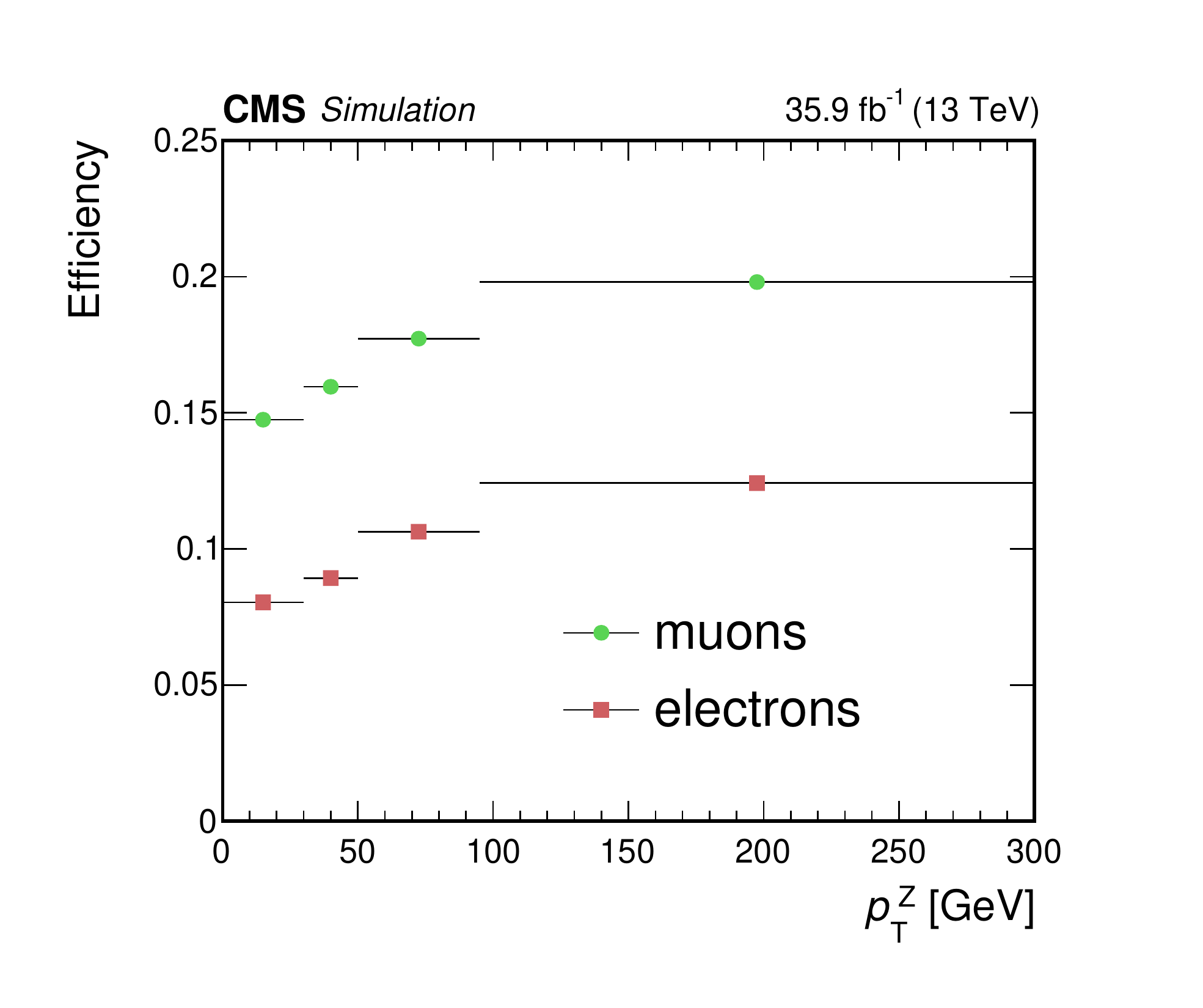}
\includegraphics[width=0.49\textwidth]{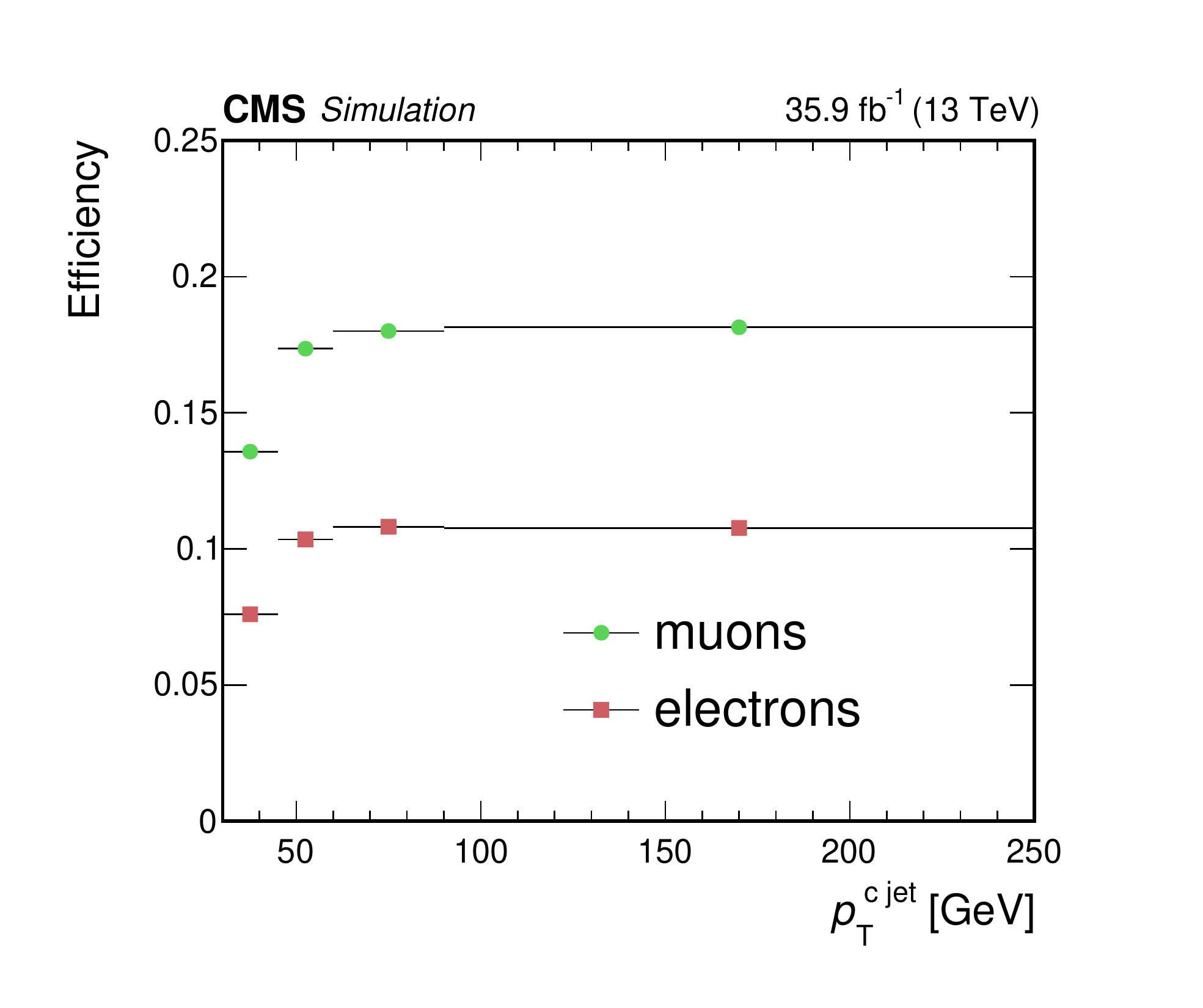} 
\caption{\label{unf:acc}Efficiency as a function of \pt. The plots
  show distributions for the electron and muon channels, as a function
  of \ptZ (left) and \ptJ (right).}
\end{figure}

Figure \ref{unf:acc} shows the efficiency (defined as the fraction of
signal events
generated in the fiducial phase space that pass all selection criteria
after reconstruction) as a function of the generator-level \Zb
or \cjet \pt for electron and muon channels, calculated using the
\mgamc (NLO) sample.
The dominant losses are due to the \PQc tagging and lepton selection efficiencies.

\section{Systematic uncertainties}
\label{sec:syst}

The systematic uncertainties are estimated by varying relevant parameters and then
repeating the unfolding procedure, recalculating the values of  
the efficiency, response matrix, and number of \zcjet and background events in
each case.
The differences observed between the unfolded distributions are assumed as the uncertainties.
The following uncertainties are included:
\begin{itemize}
  \item \textit{QCD renormalization and factorization scales:}  
    The ambiguity in the choice of QCD renormalization scale ($\mu_R$) and
    factorization scale  ($\mu_F$) leads to uncertainty in theoretical
    predictions for the DY process. This uncertainty is estimated by
    changing the values of $\mu_R$ and $\mu_F$ by factors of 0.5
    and 2 relative to  the default values, $\mu_F = \mu_R = m_{\PZ}$,
   excluding the $(0.5\mu_F,2\mu_R)$ and $(0.5\mu_R,2\mu_F)$ combinations. Largest deviations from the central values were used as uncertainty.
    \item \textit{PDF:}
The unfolding is performed with different PDF replicas and compared with the nominal distribution.

  \item \textit{\PQc tagging efficiency:}
   The effect of uncertainties in the \PQc tagging rates is estimated
   by varying tagging and mistagging scale factors for
   the different jet flavors. Scale factors for tagging \cjets,
   and mistagging \PQb jets and light jets are varied
   up and down by one standard deviation.
   The combined \PQc tagging uncertainty is then calculated as the sum
   in quadrature of these variations. The variation of scale factors
   is $\approx$15\% for light jets, and $\approx$5\% for charm and bottom jets. 
  
\item\textit{Jet energy resolution and scale:}
  Both the JES and
  JER corrections can affect jet \pt and the \SVM distributions
  used in the \SFb and \SFl measurements. The uncertainty
  resulting from JES corrections is estimated by varying the \pt- and $\eta$-dependent scale
  factors within their uncertainty (up to $\approx$4\%).
  The  JER uncertainty is estimated by varying the amount of jet \pt
  resolution degradation applied to the simulation up and down by one
  standard deviation ($\approx$10\%).  
  \item \textit{Pileup:}  
    The corresponding uncertainty is estimated by changing the total inelastic cross section by
    $\pm4.6\%$ \cite{puxs}. 
  \item \textit{Lepton identification and isolation:} 
    Uncertainties resulting from the modeling of the identification
    and isolation of muons
    and electrons are estimated by varying
    the corresponding scale factors within their uncertainties. For
    electrons the uncertainty is less than $3\%$, while for muons
    uncertainties in identification and  isolation are less than $2\%$. 
  \item \textit{Top pair production cross section:}
    The uncertainty because of the cross section used for the modeling of
    top quark pair production is estimated by varying the
    normalization of the top pair component of the background by $\pm 10\%$ \cite{ttbarxs}.
  \item \textit{Luminosity:}  
    The uncertainty is obtained by changing the luminosity value used
    to normalize the unfolded distributions by $\pm 2.5\%$ \cite{lumi}. 
  \item \textit{Statistical uncertainties in \SVM templates:}
    The uncertainty is obtained by taking into account statistical
    fluctuations in each bin of the simulated \SVM distributions, used in the fit of \SFl, \SFc and \SFb.
\end{itemize}

The uncertainties in the integral fiducial cross section from the considered sources are listed in Table \ref{tab:unc2}.

{
\begin{table}[ht]
\centering
\topcaption{Summary of the systematic uncertainties in the integral fiducial cross section arising from the various sources for electron ($\ee$) and muon ($\mumu$) channels, calculated by integrating over \ptJ or \ptZ.}
\cmsTable{
\begin{tabular}{ccccccccccc}
Channel&QCD&PDF&\PQc tag/mistag&JER&JES&Pileup&Top Pair&ID\textbackslash Iso&$\mathcal{L}$&MC stat.\\
&(\%)&(\%)&(\%)&(\%)&(\%)&(\%)&(\%)&(\%)&(\%)&(\%)\\
\hline
$\mumu,\ptJ$&5.5&0.5&4.2&3.9&4.8&1.5&0.6&1.0&2.5&4.2\\
$\mumu,\ptZ$&1.9&0.5&4.2&1.1&3.9&1.6&0.8&1.0&2.5&3.1\\
$\ee,\ptJ$&6.4&0.6&4.2&3.1&6.4&3.0&0.7&2.6&2.5&6.3\\
$\ee,\ptZ$&2.6&0.5&4.1&1.1&4.8&1.8&0.6&2.6&2.5&3.8\\
\end{tabular}
}
\label{tab:unc2}
\end{table}
}

\section{Results}

The total fiducial cross section is measured as
\begin{equation}
{\sigma_{\text{fid}}} =  \frac{{N_{\text{charm}} P_{\text{fid}}}}{{{\varepsilon}\mathcal{L}\mathcal{B}(\PZ\to \ell\ell)}},
\label{unc:finalEq}
\end{equation}
where $N_{\mathrm{charm}}$ is the integral number of measured charm events,
$P_{ \text{fid}}$ is the integral fiducial purity, $\varepsilon$ is the
integral fiducial selection efficiency, $\mathcal{L}$ is the integrated
luminosity, and $\mathcal{B}(\PZ\to\ell\ell) = 3.36 \%$ is the branching fraction of the \Zb to $\ell\ell$ with
$\ell = \Pe$ or \PGm.

The fiducial differential cross sections are obtained from the unfolded distributions as
\begin{equation}
\frac{{\rd\sigma}}{{\rd\pt}} =  \frac{{N_{i}}}{{\mathcal{L}\Delta_{i}\mathcal{B}(\PZ\to \ell\ell)}},
\label{unc:finalEq2}
\end{equation}
where $N_{i}$ is the number of events in \pt bin $i$ of the unfolded distribution and $\Delta_{i}$ is the width of the bin.

The results of the measurement of total and differential fiducial
cross sections from the electron and muon channels
are combined
by a fit using the \textsc{Convino} tool~\cite{combine}, which includes
statistical and systematic uncertainties. The uncertainties related
to the \PQc tag and mistag rates, JER, JES, pileup, luminosity, and top quark pair cross section
are assumed fully correlated between the channels, whereas uncertainties from other sources
are assumed to be uncorrelated.
The experimental systematic uncertainties are those related to
\PQc tag and mistag rates, JER, JES, identification and isolation, pileup, and luminosity.
The rest are designated as theoretical systematic uncertainties.

The total fiducial cross section value for \Zb $\pt  < 300\GeV$ equals
$405.4\pm 5.6\stat\pm 24.3\,(\text{exp}) \pm 3.7\thy\unit{pb}$, where $(\text{exp})$ and $\thy$ denote
experimental and theoretical systematic uncertainties,
respectively. This value is significantly lower than the \mgamc (NLO) predicted value of $524.9 \pm 11.7 \thy\unit{pb}$.
The theoretical systematic uncertainty includes uncertainties in QCD scale and PDF.

\begin{figure}[!htb]
    \centering
    \includegraphics[width=0.49\textwidth]{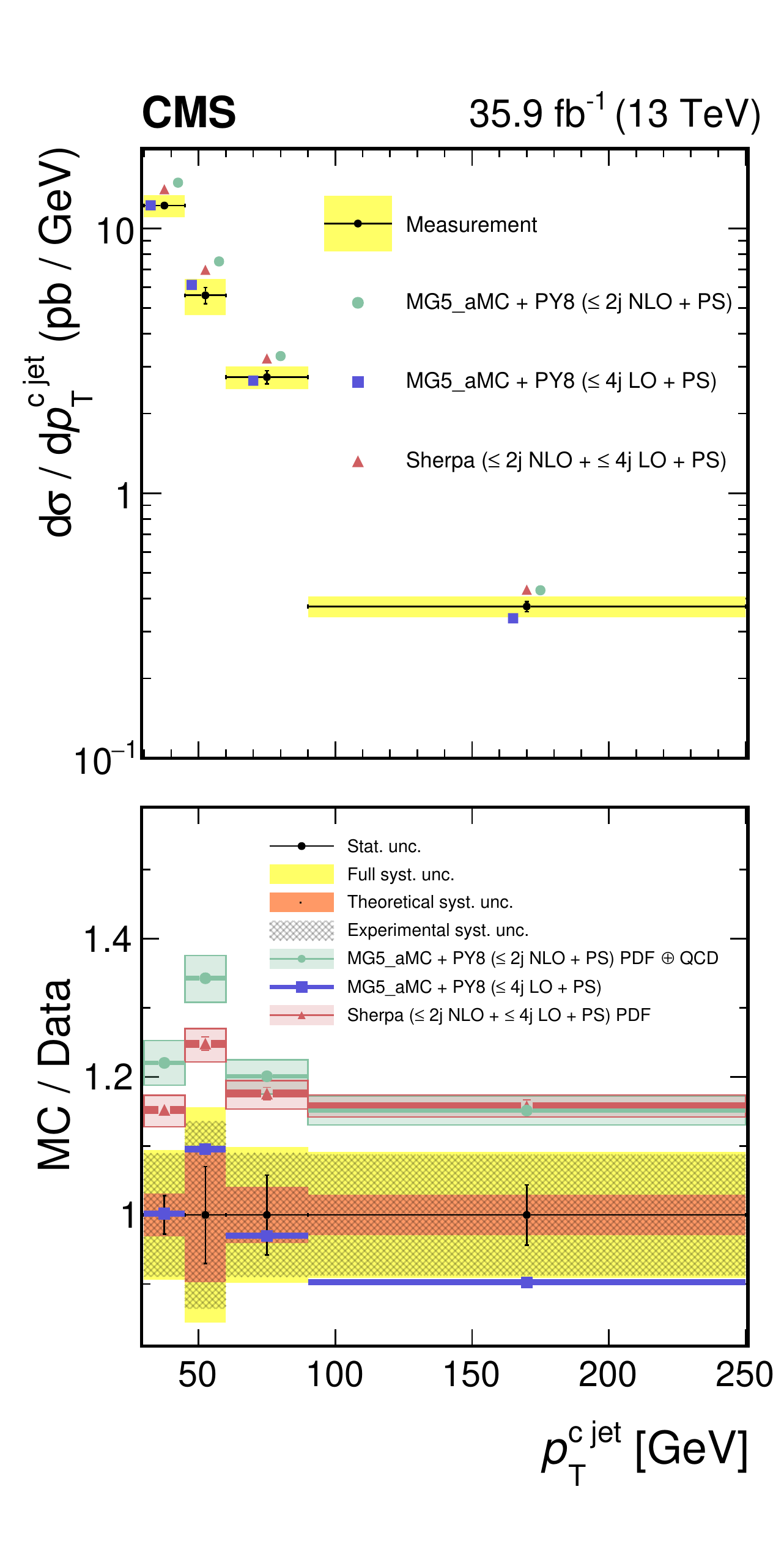}
    \includegraphics[width=0.49\textwidth]{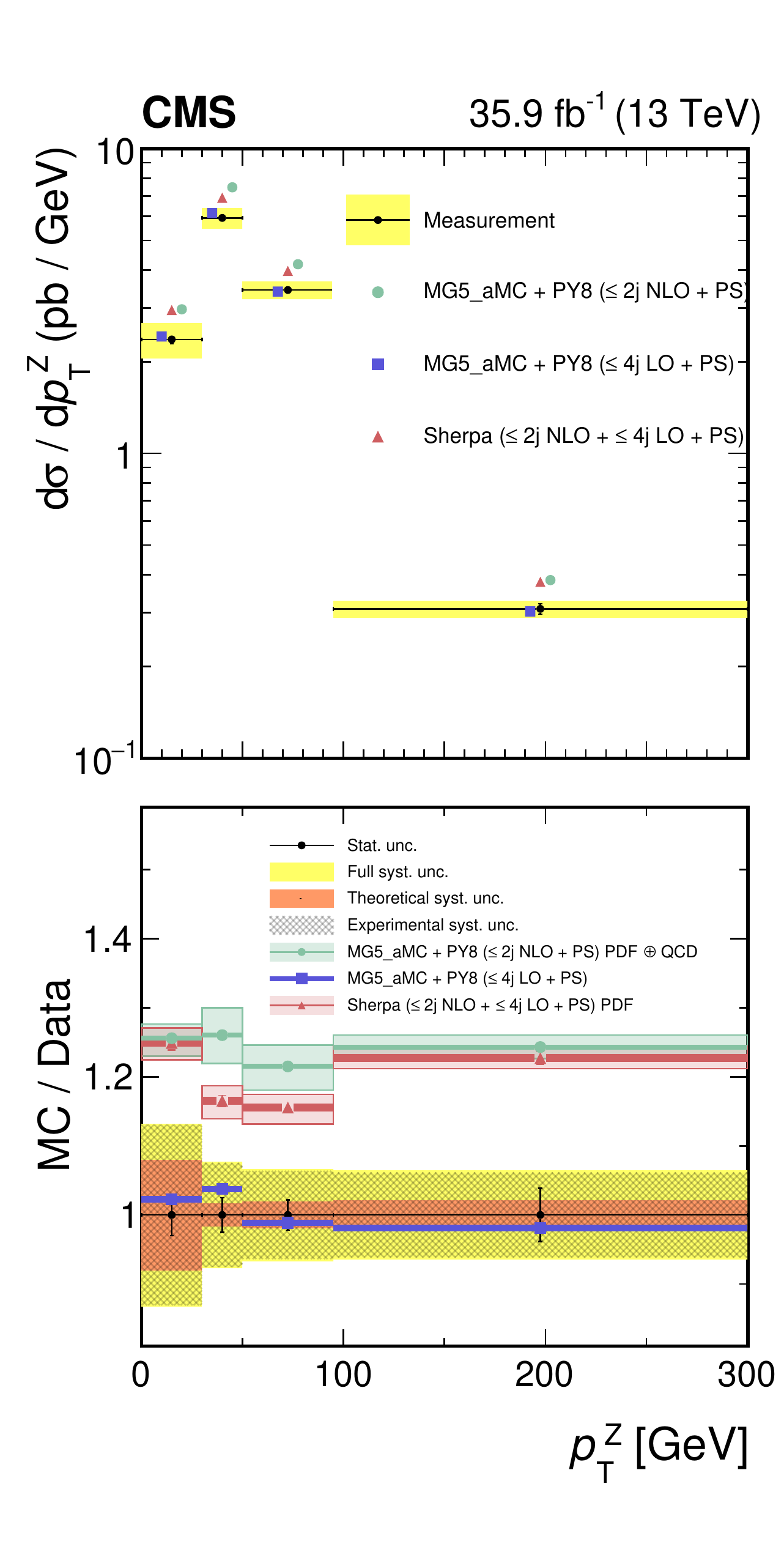}
    \caption{\label{final:plot}
      Measured fiducial differential cross sections for inclusive \zcjet production,
      ${\rd\sigma/{\rd\ptJ }}$ (left)
        and  ${\rd\sigma/{\rd\ptZ }}$ (right). Yellow band shows total systematic uncertainties. Predictions from \mgamc (LO) are shown with statistical uncertainties only. The vertical bars on the data points represent statistical uncertainties.}
    \end{figure}

The values of the cross sections as a function of \pt of the \Zb and \cjet after combining are shown in
Fig.~\ref{final:plot}. This also shows a comparison of the measured fiducial cross sections with predictions from
the generators
\mgamc (NLO),
\mgamc (LO), and \SHERPA.
The prediction from \mgamc at leading order shows good agreement with data, while both \mgamc and \SHERPA at next-to-leading order tend to overestimate the cross section.

The values of the measured differential cross sections are presented in Tables~\ref{yieldJ} and \ref{yieldZ}. 

\begin{table}[ht]
\centering
\topcaption{Measured differential cross section as a function of \ptJ for electron and muon channels and combined value. The first and second uncertainty values correspond to the statistical
 and systematic contributions, respectively.}
\begin{tabular}{cccc}
\ptJ (\GeVns)& electrons (pb/\GeVns)& muons (pb/\GeVns)& combined (pb/\GeVns)\\
\hline
30--45&11.91 $\pm$ 0.54 $\pm$ 1.50&12.34 $\pm$ 0.44 $\pm$ 1.05&12.20 $\pm$ 0.34 $\pm$ 1.15\\
45--60&5.30 $\pm$ 0.63 $\pm$ 0.92&5.73 $\pm$ 0.49 $\pm$ 0.66 &5.59 $\pm$ 0.39 $\pm$ 0.87\\
60--90&3.10 $\pm$ 0.25 $\pm$ 0.51&2.66 $\pm$ 0.19 $\pm$ 0.41&2.74 $\pm$ 0.16 $\pm$ 0.27\\
90--250&0.43 $\pm$ 0.03 $\pm$ 0.06&0.34 $\pm$ 0.02 $\pm$ 0.03&0.37 $\pm$ 0.02 $\pm$ 0.03\\
\end{tabular}
\label{yieldJ}
\end{table}

\begin{table}[ht]
\centering
\topcaption{Measured differential cross section as a function of \ptZ for electron and muon channels and combined value. The first and second uncertainty values correspond to the statistical 
and systematic contributions, respectively.}
\begin{tabular}{cccc}
\ptZ (\GeVns)& electrons (pb/\GeVns)& muons (pb/\GeVns)& combined (pb/\GeVns)\\
\hline
0--30&2.28 $\pm$ 0.13 $\pm$ 0.28 &2.40 $\pm$ 0.08 $\pm$ 0.24 &2.37 $\pm$ 0.07 $\pm$ 0.31\\
30--50&5.91 $\pm$ 0.23 $\pm$ 0.54 &5.90 $\pm$ 0.19 $\pm$ 0.46 &5.93 $\pm$ 0.15 $\pm$ 0.45\\
50--95&3.69 $\pm$ 0.13 $\pm$ 0.27&3.32 $\pm$ 0.09 $\pm$ 0.22 &3.44 $\pm$ 0.08 $\pm$ 0.23\\
95--300&0.32 $\pm$ 0.02 $\pm$ 0.03&0.30 $\pm$ 0.02 $\pm$ 0.02&0.31 $\pm$ 0.01 $\pm$ 0.02\\
\end{tabular}
\label{yieldZ}
\end{table}

\section{Summary}

The first differential cross sections for inclusive \zcjet production as functions of transverse momenta \pt of the \Zb 
and of the associated \cjet are presented for collisions at $\sqrt{s} =13\TeV$ using $35.9\fbinv$ of data collected by the CMS experiment at the CERN LHC.
 The measurements pertain to a fiducial space defined as containing a \cjet with $\pt > 30\GeV$ and pseudorapidity $\abs{\eta} < 2.4$,
 and a pair of leptons with each lepton having $\pt > 10\GeV$, $\abs{\eta} < 2.4$, and at least one with $\pt>26\GeV$, and a dilepton mass between 71 and 111\GeV.
 The main background sources correspond to \zljet, \zbjet, top quark pair, and diboson ($\PZ\PZ$, $\PZ\PW$, or $\PW\PW$) production.
 To provide a direct comparison with predictions from Monte Carlo (MC) event generators, we unfold detector effects from our measurements.

The total fiducial cross section for the \Zb with $\pt < 300\GeV$ is measured to be 
$405.4\pm 5.6\stat\pm 24.3\,(\text{exp}) \pm 3.7\thy\unit{pb}$,
 while the \MGvATNLO generator at next-to-leading order predicts $524.9 \pm 11.7 (\text{theo})\unit{pb}$ for the same fiducial region.
 The theoretical uncertainties include QCD scale variation and parton distribution function uncertainties.
 The predictions from MC event generators were compared with measurements, which are in good agreement with \MGvATNLO at leading order,
 while both \MGvATNLO and \SHERPA at next-to-leading order tend to overestimate the cross section.
 Predictions from all three generators were normalized to the cross section calculated with \FEWZ at next-to-next-to-leading order.
The prediction of inclusive {\PZ}{+}jets production at next-to-leading order is in better agreement with data than that at leading order~\cite{summary}.
 This could be an indication that the parton distribution functions overestimate the charm content.
 These results can be used to improve existing constraints on the charm quark content in the proton.

\begin{acknowledgments}
  
  We congratulate our colleagues in the CERN accelerator departments for the excellent performance of the LHC and thank the technical and administrative staffs at CERN and at other CMS institutes for their contributions to the success of the CMS effort. In addition, we gratefully acknowledge the computing centers and personnel of the Worldwide LHC Computing Grid for delivering so effectively the computing infrastructure essential to our analyses. Finally, we acknowledge the enduring support for the construction and operation of the LHC and the CMS detector provided by the following funding agencies: BMBWF and FWF (Austria); FNRS and FWO (Belgium); CNPq, CAPES, FAPERJ, FAPERGS, and FAPESP (Brazil); MES (Bulgaria); CERN; CAS, MoST, and NSFC (China); COLCIENCIAS (Colombia); MSES and CSF (Croatia); RIF (Cyprus); SENESCYT (Ecuador); MoER, ERC PUT and ERDF (Estonia); Academy of Finland, MEC, and HIP (Finland); CEA and CNRS/IN2P3 (France); BMBF, DFG, and HGF (Germany); GSRT (Greece); NKFIA (Hungary); DAE and DST (India); IPM (Iran); SFI (Ireland); INFN (Italy); MSIP and NRF (Republic of Korea); MES (Latvia); LAS (Lithuania); MOE and UM (Malaysia); BUAP, CINVESTAV, CONACYT, LNS, SEP, and UASLP-FAI (Mexico); MOS (Montenegro); MBIE (New Zealand); PAEC (Pakistan); MSHE and NSC (Poland); FCT (Portugal); JINR (Dubna); MON, RosAtom, RAS, RFBR, and NRC KI (Russia); MESTD (Serbia); SEIDI, CPAN, PCTI, and FEDER (Spain); MOSTR (Sri Lanka); Swiss Funding Agencies (Switzerland); MST (Taipei); ThEPCenter, IPST, STAR, and NSTDA (Thailand); TUBITAK and TAEK (Turkey); NASU (Ukraine); STFC (United Kingdom); DOE and NSF (USA).
  
  \hyphenation{Rachada-pisek} Individuals have received support from the Marie-Curie program and the European Research Council and Horizon 2020 Grant, contract Nos.\ 675440, 724704, 752730, and 765710 (European Union); the Leventis Foundation; the A.P.\ Sloan Foundation; the Alexander von Humboldt Foundation; the Belgian Federal Science Policy Office; the Fonds pour la Formation \`a la Recherche dans l'Industrie et dans l'Agriculture (FRIA-Belgium); the Agentschap voor Innovatie door Wetenschap en Technologie (IWT-Belgium); the F.R.S.-FNRS and FWO (Belgium) under the ``Excellence of Science -- EOS" -- be.h project n.\ 30820817; the Beijing Municipal Science \& Technology Commission, No. Z191100007219010; the Ministry of Education, Youth and Sports (MEYS) of the Czech Republic; the Deutsche Forschungsgemeinschaft (DFG) under Germany's Excellence Strategy -- EXC 2121 ``Quantum Universe" -- 390833306; the Lend\"ulet (``Momentum") Program and the J\'anos Bolyai Research Scholarship of the Hungarian Academy of Sciences, the New National Excellence Program \'UNKP, the NKFIA research grants 123842, 123959, 124845, 124850, 125105, 128713, 128786, and 129058 (Hungary); the Council of Science and Industrial Research, India; the HOMING PLUS program of the Foundation for Polish Science, cofinanced from European Union, Regional Development Fund, the Mobility Plus program of the Ministry of Science and Higher Education, the National Science Center (Poland), contracts Harmonia 2014/14/M/ST2/00428, Opus 2014/13/B/ST2/02543, 2014/15/B/ST2/03998, and 2015/19/B/ST2/02861, Sonata-bis 2012/07/E/ST2/01406; the National Priorities Research Program by Qatar National Research Fund; the Ministry of Science and Higher Education, project no. 0723-2020-0041 (Russia); the Tomsk Polytechnic University Competitiveness Enhancement Program; the Programa Estatal de Fomento de la Investigaci{\'o}n Cient{\'i}fica y T{\'e}cnica de Excelencia Mar\'{\i}a de Maeztu, grant MDM-2015-0509 and the Programa Severo Ochoa del Principado de Asturias; the Thalis and Aristeia programs cofinanced by EU-ESF and the Greek NSRF; the Rachadapisek Sompot Fund for Postdoctoral Fellowship, Chulalongkorn University and the Chulalongkorn Academic into Its 2nd Century Project Advancement Project (Thailand); the Kavli Foundation; the Nvidia Corporation; the SuperMicro Corporation; the Welch Foundation, contract C-1845; and the Weston Havens Foundation (USA).\end{acknowledgments}

\bibliography{auto_generated}   
\clearpage
\appendix
\numberwithin{figure}{section}
\section{Post-fit secondary vertex mass distributions}\label{app:svm}

\begin{figure}[!htb]
    \centering
    \includegraphics[width=0.49\textwidth]{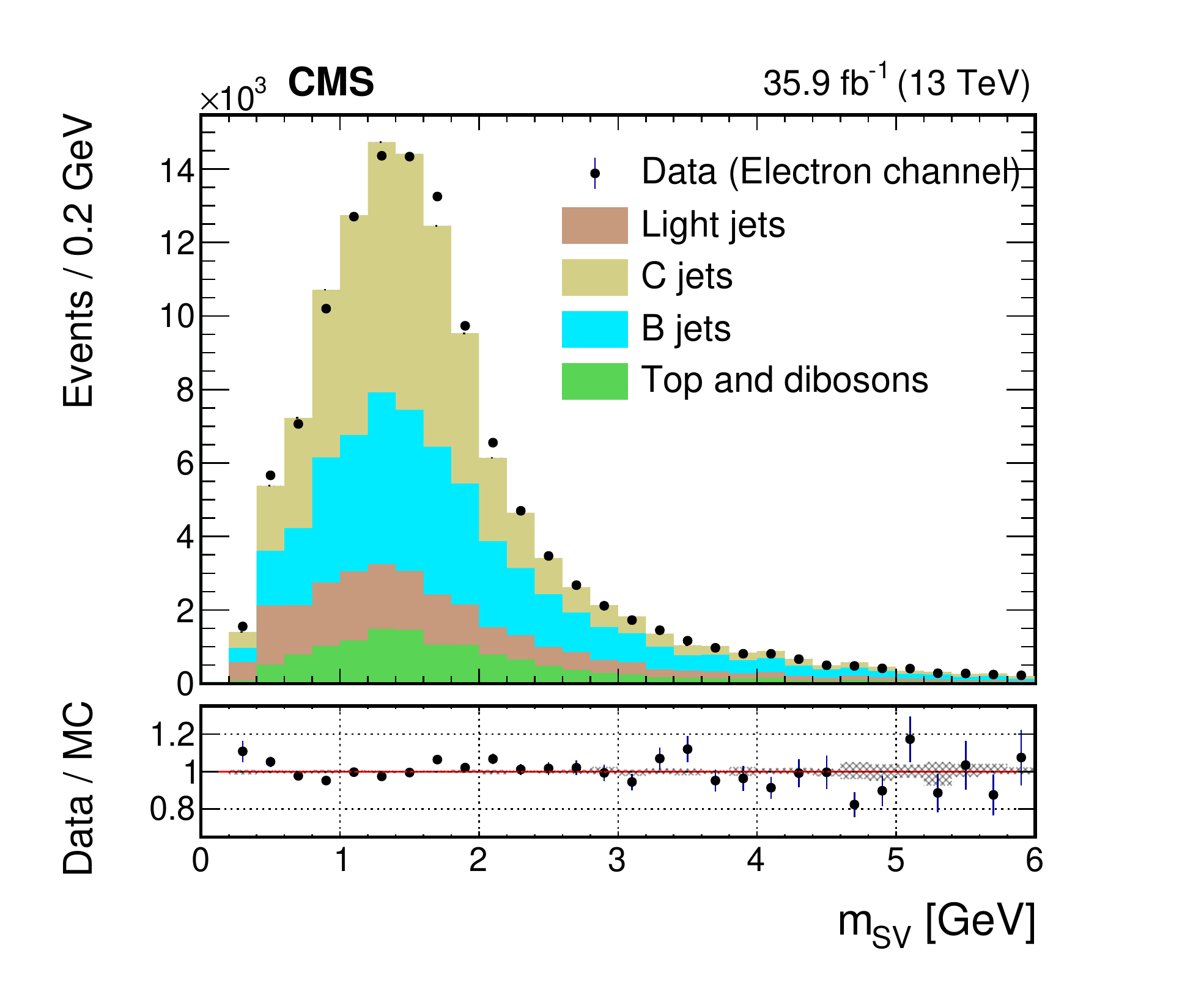}
    \includegraphics[width=0.49\textwidth]{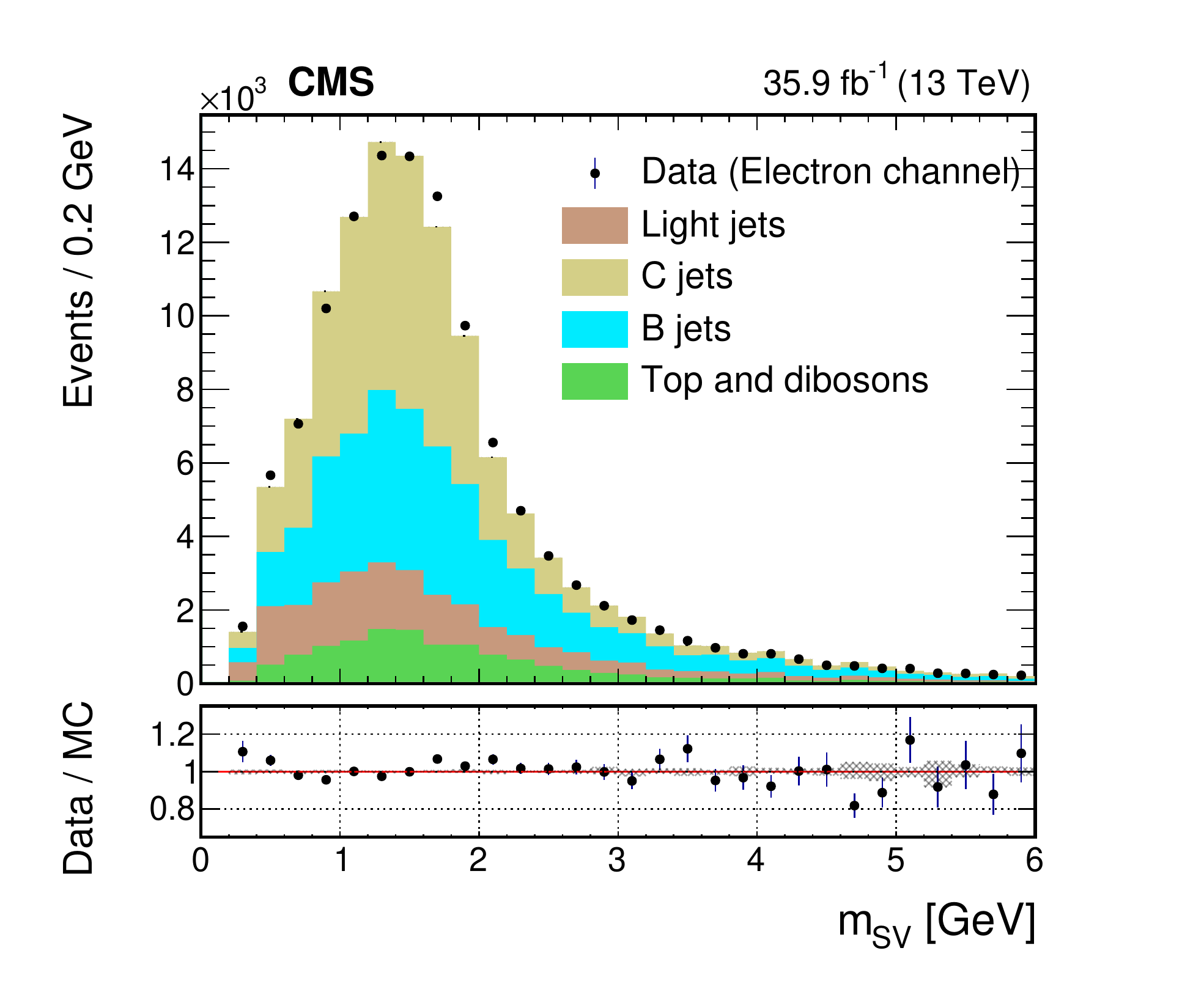}
    \caption{ \label{ctag:SVMPE} Distribution of the secondary vertex mass \SVM of the highest-{\pt}  {\PQc}-tagged central jet, for electron channel. The observed data is
      compared to the different signal and background components in
      simulation, after normalization scale factors as function of \PZ \pt (left) and {\PQc}-tagged central jet \pt (right) are applied. Bottom panels on each plot represent the data to MC ratio.}
\end{figure}

\begin{figure}[!htb]
    \centering
    \includegraphics[width=0.49\textwidth]{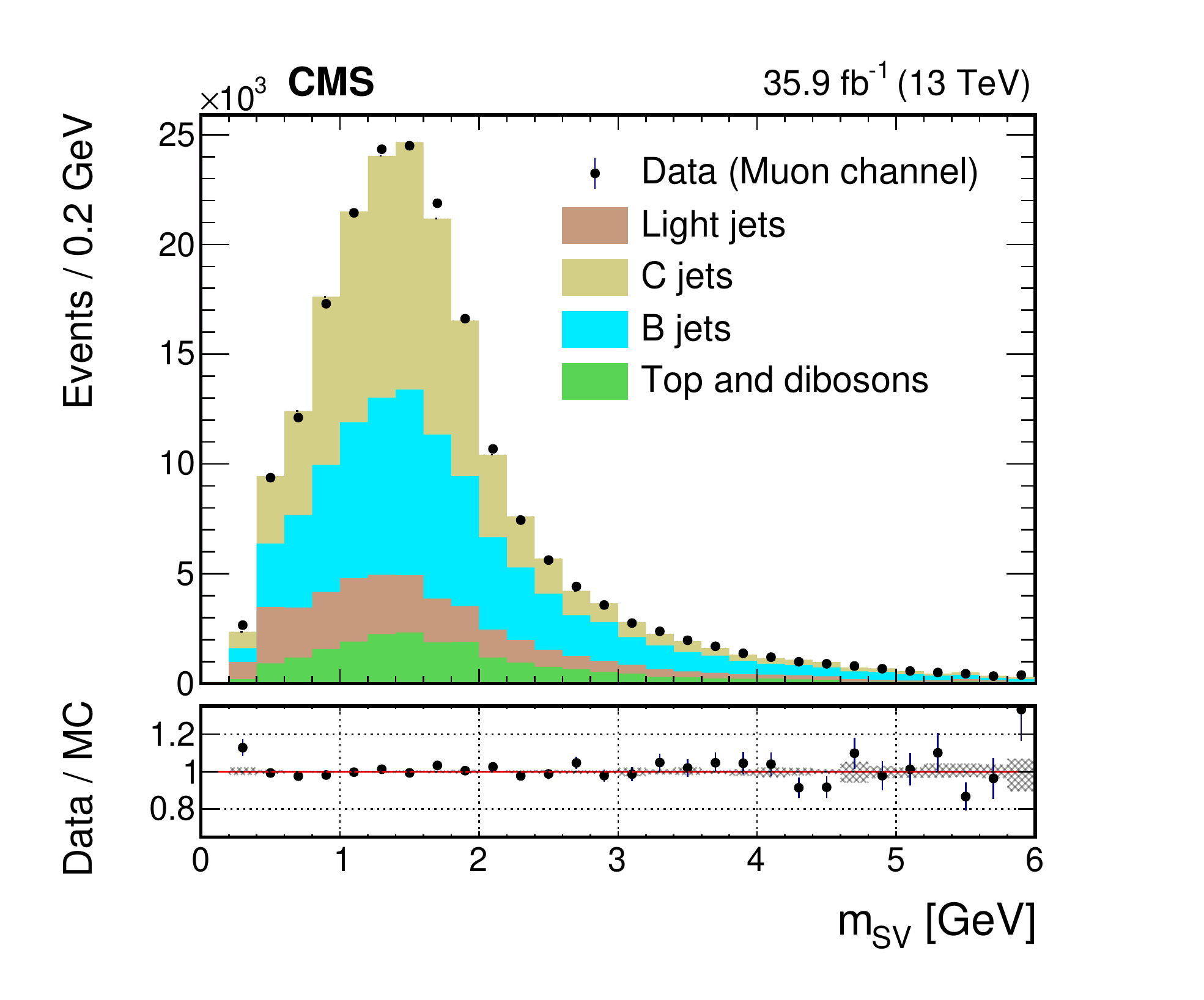}
    \includegraphics[width=0.49\textwidth]{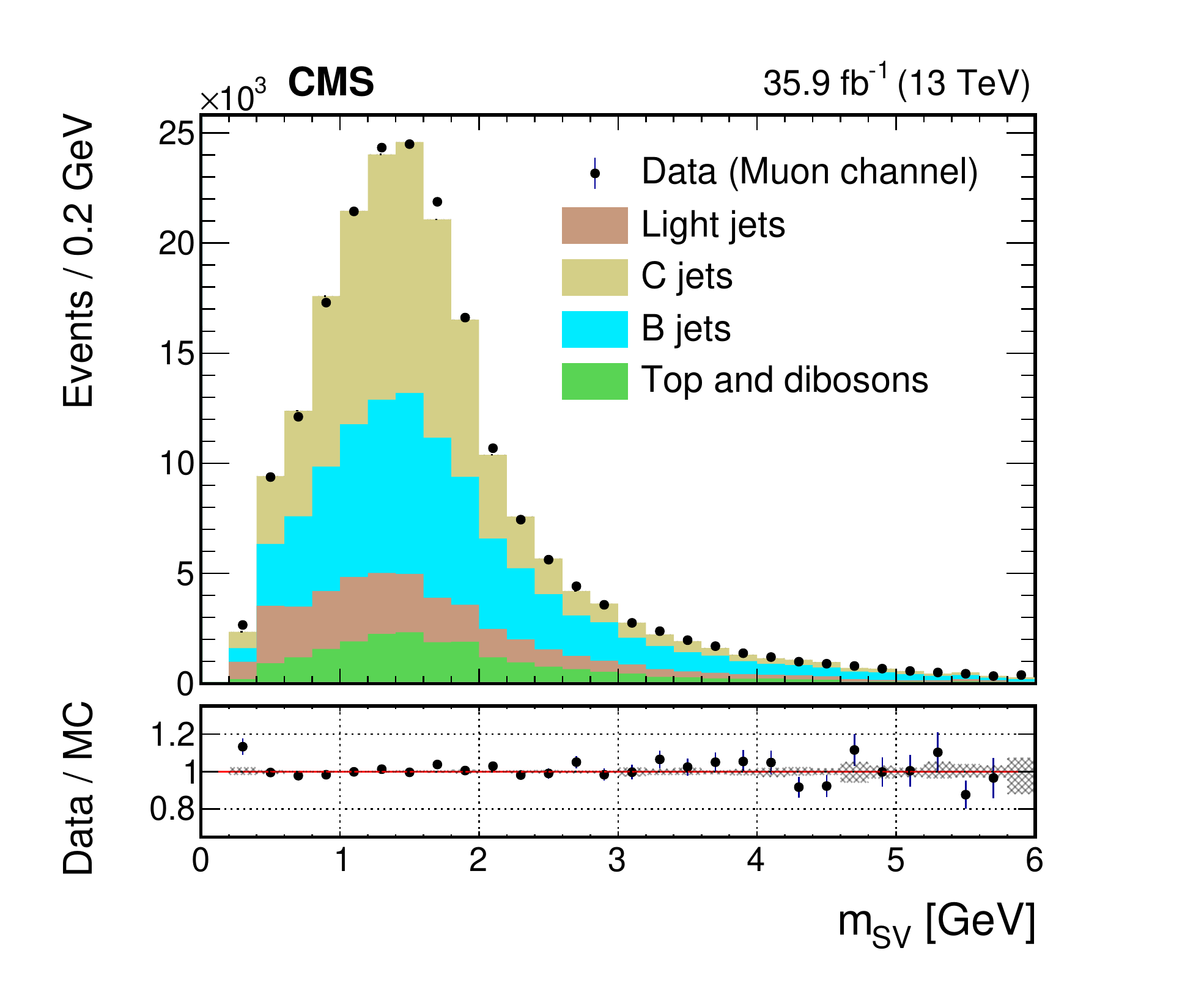}
    \caption{ \label{ctag:SVMP} Distribution of the secondary vertex mass \SVM of the highest-{\pt}  {\PQc}-tagged central jet, for muon channel. The observed data is
      compared to the different signal and background components in
      simulation, after normalization scale factors as function of \PZ \pt (left) and {\PQc}-tagged central jet \pt (right) are applied. Bottom panels on each plot represent the data to MC ratio.}
\end{figure}

Figures \ref{ctag:SVMPE} and \ref{ctag:SVMP} show post-fit secondary vertex mass distributions for electron and muon channels. The normalization scale factors 
from the fit of \SVM were applied as a function of \PZ or {\PQc}-tagged central jet \pt.

\cleardoublepage \section{The CMS Collaboration \label{app:collab}}\begin{sloppypar}\hyphenpenalty=5000\widowpenalty=500\clubpenalty=5000\input{SMP-19-011-authorlist.tex}\end{sloppypar}
\end{document}

%% file: SMP-19-011-authorlist.tex
\vskip\cmsinstskip
\textbf{Yerevan Physics Institute, Yerevan, Armenia}\\*[0pt]
A.M.~Sirunyan$^{\textrm{\dag}}$, A.~Tumasyan
\vskip\cmsinstskip
\textbf{Institut f\"{u}r Hochenergiephysik, Wien, Austria}\\*[0pt]
W.~Adam, T.~Bergauer, M.~Dragicevic, J.~Er\"{o}, A.~Escalante~Del~Valle, R.~Fr\"{u}hwirth\cmsAuthorMark{1}, M.~Jeitler\cmsAuthorMark{1}, N.~Krammer, L.~Lechner, D.~Liko, T.~Madlener, I.~Mikulec, F.M.~Pitters, N.~Rad, J.~Schieck\cmsAuthorMark{1}, R.~Sch\"{o}fbeck, M.~Spanring, S.~Templ, W.~Waltenberger, C.-E.~Wulz\cmsAuthorMark{1}, M.~Zarucki
% ...
\vskip\cmsinstskip
\dag: Deceased\\
1:  Also at Vienna University of Technology, Vienna, Austria\\
% ...
